\documentclass[useAMS,usenatbib]{mnras}
\usepackage{graphicx,epsfig,longtable}
\usepackage{epstopdf}
\usepackage{color}
\usepackage{amsmath}
\usepackage{tabularx} 

\newcommand{\Msun}{$M_{\sun}$}

\title {Dynamical ejecta of neutron star mergers with nucleonic weak processes I: Nucleosynthesis}
\author[I.~Kullmann et al.]
  {I. Kullmann$^1$, S.~Goriely$^1$, O.~Just$^2$, R. Ardevol-Pulpillo$^3$, A. Bauswein$^2$,   \newauthor H.-T. Janka$^3$ \\ 
  $^1$Institut d'Astronomie et d'Astrophysique, CP-226, Universit\'e Libre de Bruxelles, 1050 Brussels, Belgium\\
  $^2$GSI Helmholtzzentrum f\"ur Schwerionenforschung, Planckstrasse 1, 64291 Darmstadt, Germany\\
  $^3$Max-Planck-Institut f\"ur Astrophysik, Postfach 1317, 85741 Garching, Germany 
 }
\date{Released 2021 Xxxxx XX}

\pagerange{\pageref{firstpage}--\pageref{lastpage}} \pubyear{2014}

\def\ga{\,\,\raise0.14em\hbox{$>$}\kern-0.76em\lower0.28em\hbox
{$\sim$}\,\,}
\def\la{\,\,\raise0.14em\hbox{$<$}\kern-0.76em\lower0.28em\hbox
{$\sim$}\,\,}
\def\Msun{$M_{\odot}$}

\begin{document}

\label{firstpage}

\maketitle

\begin{abstract}
We present a coherent study of the impact of neutrino interactions on the r-process element nucleosynthesis and the heating rate produced by the radioactive elements synthesised in the dynamical ejecta of neutron star-neutron star (NS-NS) mergers. 
We have studied the material ejected from four NS-NS merger systems based on hydrodynamical simulations which handle neutrino effects in an elaborate way by including neutrino equilibration with matter in optically thick regions and re-absorption in optically thin regions. We find that the neutron richness of the dynamical ejecta is significantly affected by the neutrinos emitted by the post-merger remnant, in particular when compared to a case neglecting all neutrino interactions. Our nucleosynthesis results show that a solar-like distribution of r-process elements with mass numbers $A \ga 90$ is produced, including a significant enrichment in Sr and a reduced production of actinides compared to simulations without inclusion of the nucleonic weak processes.
 The composition of the dynamically ejected matter as well as the corresponding rate of radioactive decay heating are found to be rather independent of the system mass asymmetry and the adopted equation of state. This approximate degeneracy in abundance pattern and heating rates can be favourable for extracting the ejecta properties from kilonova observations, at least if the dynamical component dominates the overall ejecta. Part II of this work will study the light curve produced by the dynamical ejecta of our four NS merger models.
\end{abstract}


\section{Introduction}

A long-standing science puzzle has been to explain the origin of the heaviest elements in the Universe, and more particularly the production of the elements heavier than iron up to uranium. One of the nucleosynthesis processes called for to explain the origin of these heavy elements is the rapid neutron-capture process or r-process \citep{burbidge1957,cameron1957,arnould2020,arnould2007}.  It can account for the stable (and some long-lived radioactive) neutron-rich nuclides observed in stars of various metallicities, as well as in the Solar system, {\it i.e.} about half of the elements heavier than Fe. However, for decades there existed no direct evidence of where in the Universe this process could occur \citep{arnould2007,cowan2021}. 

The r-process remains the most complex nucleosynthetic process to model from the astrophysics as well as nuclear-physics points of view. Early studies of core-collapse supernovae have raised a lot of excitement about the so-called neutrino-driven wind environment \citep{qian1996,hoffman1997,Arcones2007}.
However, until now a successful r-process cannot be obtained {\it ab initio} without tuning the relevant parameters (neutron excess, entropy, expansion timescale) in a way that is not supported by increasingly more sophisticated existing models \citep{witti1994,takahashi1994,roberts2010,wanajo2011,janka2012,wanajo2018c}.
Although these scenarios remain promising, especially in view of their potential to contribute to the galactic enrichment significantly, they remain affected by large uncertainties associated mainly with the still complex mechanism responsible for the  supernova explosion and the persistent difficulties to obtain suitable r-process conditions in self-consistent dynamical explosion and neutron-star (NS) cooling models \citep{janka2012,hudepohl2010,fischer2010,Janka2016,Muller2017,Bollig2017,Bolling2021,Pinedo2012}. 
Moreover, a subclass of core-collapse supernovae, the so-called jet-driven or jet-associated collapsars corresponding to the death of rapidly rotating, massive stars, some of which might also be the origin of observed long $\gamma$-ray bursts \citep{Woosley1993,macfadyen1999}, could represent another, promising r-process site. In these systems, r-process viable conditions may be found in magnetically driven jets \citep{cameron2003,winteler2012,Nishimura2015,Mosta2018,Reichert2021,Grimmett2020} or in the ejecta launched from the central black-hole accretion disk \citep{Cameron2001,Pruet2004, Surman2004, Siegel2019, miller2019a,just2021}.

Special attention has also been paid to an alternative site, namely NS mergers following the initial study of the decompression of cold neutronized matter by \cite{lattimer1974,lattimer1977,meyer1989}, who first proposed that merging binary NSs could eject neutron-rich material and be a suitable site where the r-process nucleosynthesis could take place. Recent studies including hydrodynamical simulations of NS merger systems have established that a significant amount of material ($\sim10^{-3}-10^{-1}$ \Msun) can be ejected during such events \citep[e.g.][]{rosswog1999, ruffert2001, oechslin2007, hotokezaka2013, bauswein2013, just2015,radice2018b,foucart2016}. 
It was found that the neutron-rich conditions in the ejected material enable a robust r-process, producing a solar like abundance distribution up to the 3rd r-process peak and the actinides \citep[e.g.][]{freiburghaus1999, goriely2011, korobkin2012, wanajo2014, just2015, radice2018b}. 

The r-nuclide enrichment is predicted to originate from both the dynamical (prompt) material expelled during the NS-NS or NS-BH merger phase and from the outflows generated during the evolution of the merger remnant, which is a massive NS or a BH surrounded by an accretion torus. However, the detailed composition of each ejecta component, and therefore its role for the galactic chemical evolution, remains uncertain at this point, which is mainly owing to the computational difficulties associated with the treatment of neutrino interactions and details of the ejection mechanism. The impact of neutrinos is particularly challenging to assess for the dynamical ejecta, first because the phase of the merger requires a three-dimensional, general relativistic description and cannot be approximated by an axisymmetric evolution, and second, the evolution during the merger phase is extremely violent and characterized by short dynamical timescales and strong shocks. Early studies \citep[e.g.][]{goriely2011,korobkin2012,mendoza-temis2015} found very neutron-rich ($Y_e\la 0.1$) dynamical outflows and a significant solar-like production of $A>140$ elements. However, the impact of weak processes on the ejecta was either neglected in those simulations or was relatively minor because of dominant tidal ejection \citep{korobkin2012} and, under such an approximation, the r-process nucleosynthesis yields were shown to be rather insensitive to the initial astrophysical conditions, such as the masses of the binary NS or the nuclear equation of state (EoS), due to fission recycling \citep{goriely2011}.

This picture changed when weak interactions started to be included in both the hydrodynamical simulations as well as the nucleosynthesis calculations \citep{wanajo2014, goriely2015a, martin2018, radice2016, foucart2016}. The weak reactions that are most important are the $\beta$-interactions of electron neutrinos ($\nu_e$) with free neutrons ($n$) and electron antineutrinos ($\bar{\nu}_e$) with free protons ($p$) as well as their inverse reactions,
\begin{equation}
\begin{aligned}
 & \nu_e+ n \rightleftharpoons p + e^- , \label{eq:betareac}  \\
 & \bar{\nu}_e + p \rightleftharpoons n + e^+  .
\end{aligned}
\end{equation}
In studies incorporating the above neutrino emission/absorption reactions the neutron richness of the dynamical ejecta was shown to be substantially altered by the neutrinos emitted by the post-merger remnant, driving the electron fraction up to $\sim0.3-0.4$ and even $\sim0.5$ in the polar regions. However, the exact impact of these nucleonic weak processes in NS mergers still remains unclear for several reasons. Firstly, there is a large spread in the applied approximations for the neutrino treatment in the literature or simply many studies do not take neutrino absorption into account \citep[e.g.][]{wanajo2014, palenzuela2015,lehner2016,sekiguchi2015,foucart2016,bovard2017,radice2016,martin2018}. Secondly, the quantitative effects of weak reactions may depend on the details of the merger dynamics and the shock heating strength during the merging, and therefore on the adopted EoS \citep{sekiguchi2015} and possibly, to some extent, on numerics. In addition, flavour conversions of the neutrino species can also affect the neutron richness, hence the nucleosynthesis in the merger ejecta \citep{wu2017,george2020}. 

The decay of freshly produced radioactive r-process elements can power a ``kilonova'', which is potentially observable in the optical electromagnetic spectrum in the aftermath of a NS merger event \citep[e.g.][]{li1998, roberts2011, metzger2010, goriely2011, barnes2013,Kulkarni2005,Tanaka2013}. Indeed, such an electromagnetic counterpart, AT2017gfo \citep{abbott2017d} was recently observed for the first time accompanying the gravitational wave signal, GW170817 \citep[e.g.][]{abbott2017c}, from two merging NSs. Fits to the observed kilonova light curve provided a strong indication that r-process elements have been synthesized in this event \citep[e.g.][]{kasen2017}. Recently, \cite{watson2019} also reported the spectroscopic identification of strontium, providing the first direct evidence that NS mergers eject material enriched in heavy elements. The fact that strontium was present implies that weak interactions may play a prominent role to shape the composition of the ejecta, since weak interactions alter the initial composition towards higher $Y_e$ values.

Models of the kilonova observed in the aftermath of GW170817 have indicated the presence of (at least) two kilonova components, the so-called ``blue'' and ``red''  components\footnote{Note that the ``blue" and ``red" kilonova components did not exhibit spectral peaks in the blue optical and red optical bands, but rather in the red optical and near infrared bands, respectively.},  which are presumably associated with ejecta material having $A<140$ and $A>140$, respectively \citep[e.g.][]{kasen2017}.  However, given the complexity of modelling the merger and its remnant the identification of these kilonova components with the basic ejecta components (dynamical vs. winds from a hypermassive NS or from a remnant BH-disk) is not yet unambiguous. In particular, the role of the dynamical ejecta, e.g. whether they contribute more to the blue or more to the red kilonova component, is not finally settled \citep[e.g.][]{kasen2017,Kawaguchi2018}.

This work will contribute to the question of whether a more accurate description of neutrino interactions can significantly affect the r-process abundance yields in NS mergers.  
To this end, r-process nucleosynthesis network calculations will be performed on the full set of ejected trajectories from four relativistic hydrodynamical simulations using the so-called improved leakage-equilibration-absorption scheme (ILEAS) for neutrino transport  \citep{ardevol-pulpillo2019}. ILEAS takes into account neutrino equilibration with matter in optically thick regions, uses an improved prescription of the diffusion timescale that is consistent with the diffusion law, and describes the transition to free streaming via a flux-limiter. Moreover, ILEAS includes re-absorption in optically thin regions. Hence, ILEAS should provide a fairly realistic picture of the impact of neutrino interactions on the early ejecta in NS mergers.
The present study is divided into two papers. Part I, the present paper, focuses on the r-process nucleosynthesis and the subsequent heating rate produced in the dynamical ejecta, while Part II \citep{just2021b} studies consistently ({\it i.e.} on the basis of our four NS merger models and their determined nucleosynthesis) the resulting electromagnetic signal expected from the dynamical ejecta. 
Since we will focus here on the dynamical ejecta only, neglecting the possible contribution from the hypermassive NS or the remnant BH-disk winds, no attempt is made to compare our results with observations, neither of kilonovae nor from spectroscopy of r-process-rich stars. This aspect will be studied in another forthcoming paper.
 
In Sect.~\ref{sect_weak}, we introduce our NS merger models and their basic properties. We discuss the ejecta electron fraction and show that the neutron richness is significantly affected by the presence of neutrinos, in particular when compared to cases neglecting neutrino absorption and emission. The results of the nucleosynthesis calculations are discussed in Sect.~\ref{sec_nuc_q}, describing both the composition and the radioactive decay heat of the ejected material. Special attention is paid to the angular and velocity dependence of our nucleosynthesis results which are analysed in detail in Sect.~\ref{sec_ang_vc}. A comparison of our results to other simulations including nucleonic weak interactions, similar NS mass configurations and EoSs is performed in Sect.~\ref{sect_comp}.
Final remarks and conclusions are given in Sect. \ref{sect_concl}.

\section{NS merger models}
\label{sect_weak}

In the present study we analyse the composition of material dynamically ejected from two binary NS merger systems using two different EoSs,  giving in total four models. 
The hydrodynamical models are based on the work of \citet{ardevol-pulpillo2019} that includes simulations of symmetric binary merger systems consisting of 1.35-1.35\Msun\ NSs, which have been extended to include asymmetric systems consisting of 1.25-1.45\Msun\ NSs.  
In these simulations, the recent ILEAS method is coupled to a relativistic smoothed-particle-hydrodynamics (SPH) code that assumes a conformally flat spatial metric to reduce the complexity of the Einstein equations. 
As shown for instance in \citet{Ciolfi20}, the impact of magnetic fields on the properties of the dynamical ejecta can be neglected.
ILEAS is designed to follow the neutrino effects in the hot and dense astrophysical plasma \citep{ardevol-pulpillo2019}. This approximation improves the lepton number and energy losses of traditional leakage descriptions by a novel prescription of the diffusion time-scale based on a detailed energy integral of the flux-limited diffusion equation. The leakage module is also supplemented by a neutrino-equilibration treatment that ensures the proper evolution of the total lepton number and medium plus neutrino energies as well as neutrino-pressure effects in the neutrino-trapping domain. In addition,  a simple and straightforwardly applicable ray-tracing algorithm is used for including re-absorption of escaping neutrinos especially in the decoupling layer and during the transition to semitransparent conditions.  All details can be found in \cite{ardevol-pulpillo2019} where it was shown in particular that ILEAS can satisfactorily reproduce local losses and re-absorption of neutrinos as found in more sophisticated transport calculations on the level of 10\%\footnote{The tests in \cite{ardevol-pulpillo2019} were performed for a spherically symmetric proto-NS configuration and a BH-torus. The accuracy for the merger case is likely similar though not exactly known due to the lack of detailed reference solutions. }.
At the start of the hydrodynamical simulation, a few orbits before the merging of the two NSs, equilibrium is assumed to hold between electrons, positrons and neutrinos. 
The Cartesian grid covering the NS in which the neutrino interactions are calculated expands 100~km in all six directions from the centre of mass of the system, with a resolution of 0.738~km\footnote{
Although detailed resolution tests have yet to be conducted, the rather modest resolution of 0.738~km can be justified by three main arguments. First, the density gradients at the neutron star surfaces are steep only prior to the collision of the two stars, {\it i.e.} at times when the neutrino emission is still low. Steep density gradients develop again at late post-merger times after considerable neutrino cooling of a stable merger remnant. Such a late evolution phase, however, is not reached in our present set of merger models. Second, our neutrino treatment is not based on numerically determined gradients but employs space integrals that are usually less sensitive to fine spatial structures. Third, the possibility of resolving steep density gradients or local density variations on small scales is essentially set by the numerical limits from SPH hydrodynamics rather than from the neutrino treatment.}. 
After the merger, as the remnant torus expands, the grid size is progressively increased keeping the whole remnant covered at all times and the same resolution (cell size) throughout the simulation \citep{ardevol-pulpillo2019}.
These relativistic hydrodynamic simulations also provide the temperature evolution to which no post-processing is applied \cite[in contrast to][]{goriely2011} to retain consistency between the neutrino properties and the other thermodynamical properties.
In these simulations, either the temperature-dependent DD2 \citep{hempel2010,typel2010} or SFHo EoS \citep{steiner2013} is adopted.
For clarity,  from now on the DD2 1.35-1.35\Msun\ and 1.25-1.45\Msun\ merger models will be referred to by DD2-135135 and DD2-125145, respectively,  and similarly the SFHo 1.35-1.35\Msun\ and 1.25-1.45\Msun\ models as SFHo-135135 and SFHo-125145, respectively. 

\begin{figure*}
\begin{center}
\includegraphics[width=\textwidth]{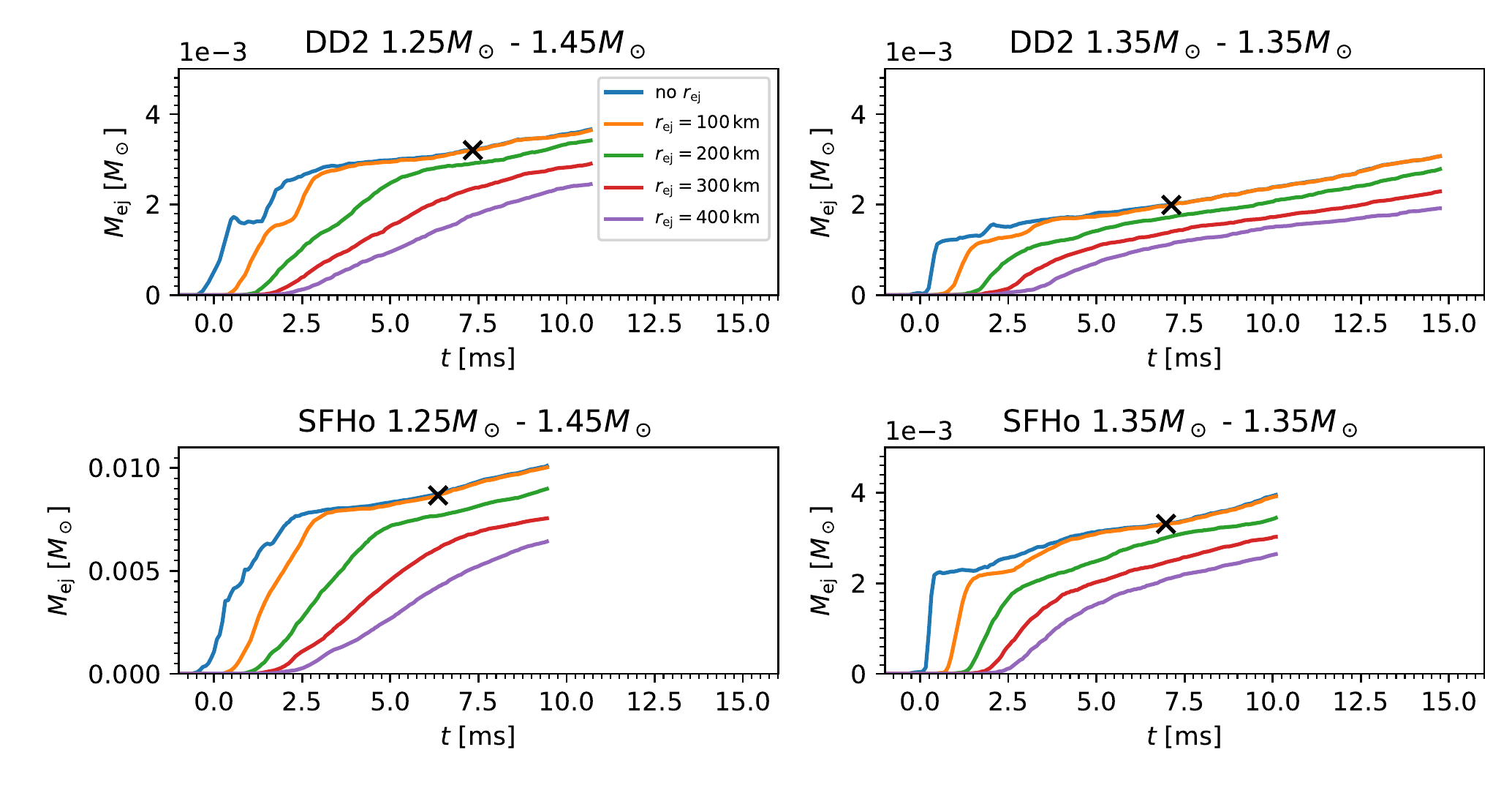}
\caption{(Color online). 
Time evolution of the dynamically ejected masses for our four merger
models as measured by the criterion $\varepsilon_{\rm stationary}>0$ applied beyond radii of $r_{\rm ej}=0$, 100~km, 200~km, 300~km, 400~km. The $t=0$ reference time corresponds to the phase of the merger when the remnant is maximally compressed about 1~ms after the stars first touched. The fact that lines for different values of $r_{\rm ej}$ are similar to each other indicates that material flagged
unbound at low radii keeps expanding and truly ends up as ejected
material. The crosses at time $t=t_{\rm ej}\sim 7$~ms for lines with $r_{\rm ej}=100$~km correspond to the ejecta material that is adopted for nucleosynthesis calculations in this study (see also Table~\ref{tab:prop}).
}
\label{fig_mejec}
\end{center}
\end{figure*}

For the four models, the number of unbound mass elements are 783,  1263,  1290 and 4398 with a total ejected mass of $2\times 10^{-3}$\Msun, $3.2\times10^{-3}$\Msun, $3.3\times10^{-3}$\Msun\ and $8.6\times10^{-3}$\Msun\ for  DD2-135135,  DD2-125145, SFHo-135135 and SFHo-125145, respectively. 
It should, however, be stressed that defining the exact total amount of mass dynamically ejected from the system is far from being an easy task \citep[see e.g.][]{Foucart2021}, since different criteria for gravitationally unbound conditions can be considered, and furthermore, the adopted criterion can be applied at different times or radii. 
We employ the positive stationary energy criterion  ($\varepsilon_{\rm stationary}>0$) stating that a mass element becomes unbound if at a chosen time $t_{\rm ej}$ and beyond a certain radius $r_{\rm ej}$, the kinetic plus thermal energy exceeds the gravitational one \citep[modulo relativistic corrections, see, e.g., ][for a detailed derivation of this criterion]{oechslin2007}.
In Fig.~\ref{fig_mejec} we compare for each model the growth of $M_{\rm ej}$ with time, where $M_{\rm ej}(t)$ is the mass of all particles that fulfill $\varepsilon_{\rm stationary}>0$ and $r >
r_{\rm ej}$, and $r_{\rm ej}$ is varied between 0, 100~km, 200~km, 300~km, and 400~km. Independent of the chosen value of $r_{\rm ej}$, the total ejected mass is found to grow with time and not to saturate at the end of the simulations.  In order to select ejecta particles for our nucleosynthesis calculations, an ejection
radius of 100~km and time of 7~ms are adopted, as shown by the crosses in Fig.~\ref{fig_mejec}.
A finite value of $r_{\rm ej}$ is chosen to ensure the correct
identification of expanding material, however, as visible in Fig.~\ref{fig_mejec}, the choice of $r_{\rm ej}=100$~km is equivalent to applying no radius criterion at all. The similar, nearly parallel, line shape for different values of $r_{\rm ej}$ indicates the robustness of using a low value of $r_{\rm ej}$ in the sense that a larger value of $r_{\rm ej}$ would count the same ejecta material but just at a correspondingly later time $t_{\rm ej}$ (the time needed for material to reach that particular  $r_{\rm ej}$). We indeed verified that, for model DD2-135135, nearly the same mass elements are counted when using $t_{\rm ej}=10$~ms and $r_{\rm ej}=220$~km instead of 7~ms and 100~km, respectively. 
Note that the time $t_{\rm ej}$ is chosen in the middle of the shallow, nearly linear growth since at later times the ejecta are more likely to be affected by magneto-hydrodynamical and secular effects in the merger remnant. The exact physical reasons behind such a continuous increase in the mass loss has not been determined, but it could be associated with various mechanisms, such as a growing 
$m=1$ mode \citep{nedora2019}, neutrino-driven winds, or viscous effects and angular momentum redistribution in the merger remnant.
The ejected masses for the different models are summarised in Table~\ref{tab:prop}. See Sect. ~\ref{sect_comp} for a comparison to other works.

For each of these relativistic hydrodynamical simulations, a post-processing reaction network is used to calculate the composition and radioactive decay heat of the ejected material.
The nucleosynthesis calculation starts as soon as the temperature drops below  $T=10^{10}$~K and the density is below the neutron-drip density $\rho_\mathrm{drip}\simeq 4.2\times 10^{11}$\,g\,cm$^{-3}$, at which the initial abundances of heavy nuclei are determined by nuclear statistical equilibrium at the given electron fraction, density and temperature. 
The density at which the full reaction network is initiated is referred to as $\rho_\mathrm{net}$.  
To take advantage of the complete ILEAS calculation,  the initial $Y_e$ adopted by the network calculation corresponds to the value estimated by the hydrodynamical model at the end of the simulation, {\it i.e.} some 10~ms after the plunge. Note that the evolution of the $Y_e$ rapidly reaches a plateau before the end of the hydrodynamical simulation and before the beginning of the network calculation, for which reason the difference in $Y_e$ at $\rho_\mathrm{net}$ and at the last hydrodynamical time step is found to be smaller than 0.01. Only a handful of trajectories have a $Y_e$ difference larger than 0.01, though always smaller than 0.05. 

For the first 10\,ms after the plunge the density history is consistently followed by the numerical 
simulation. Afterwards the ejected matter is assumed to expand freely with constant velocity. The radii of the ejecta clumps thus grow linearly with time $t$ and consequently their densities drop like $1/t^3$. As soon as the full reaction network is initiated at $\rho=\rho_\mathrm{net}$, the temperature evolution is determined on the basis of the laws of thermodynamics, allowing for possible nuclear heating through $\beta$-decays, fission, and $\alpha$-decays \citep{meyer1989}.
During this extrapolation,  the weak interactions of nucleons (Eqs.~\ref{eq:betareac}) are neglected since the hydrodynamical simulation shows that the $Y_e$ has reached a plateau and is not affected anymore by these weak interactions.

\begin{figure}
\includegraphics[scale=0.33]{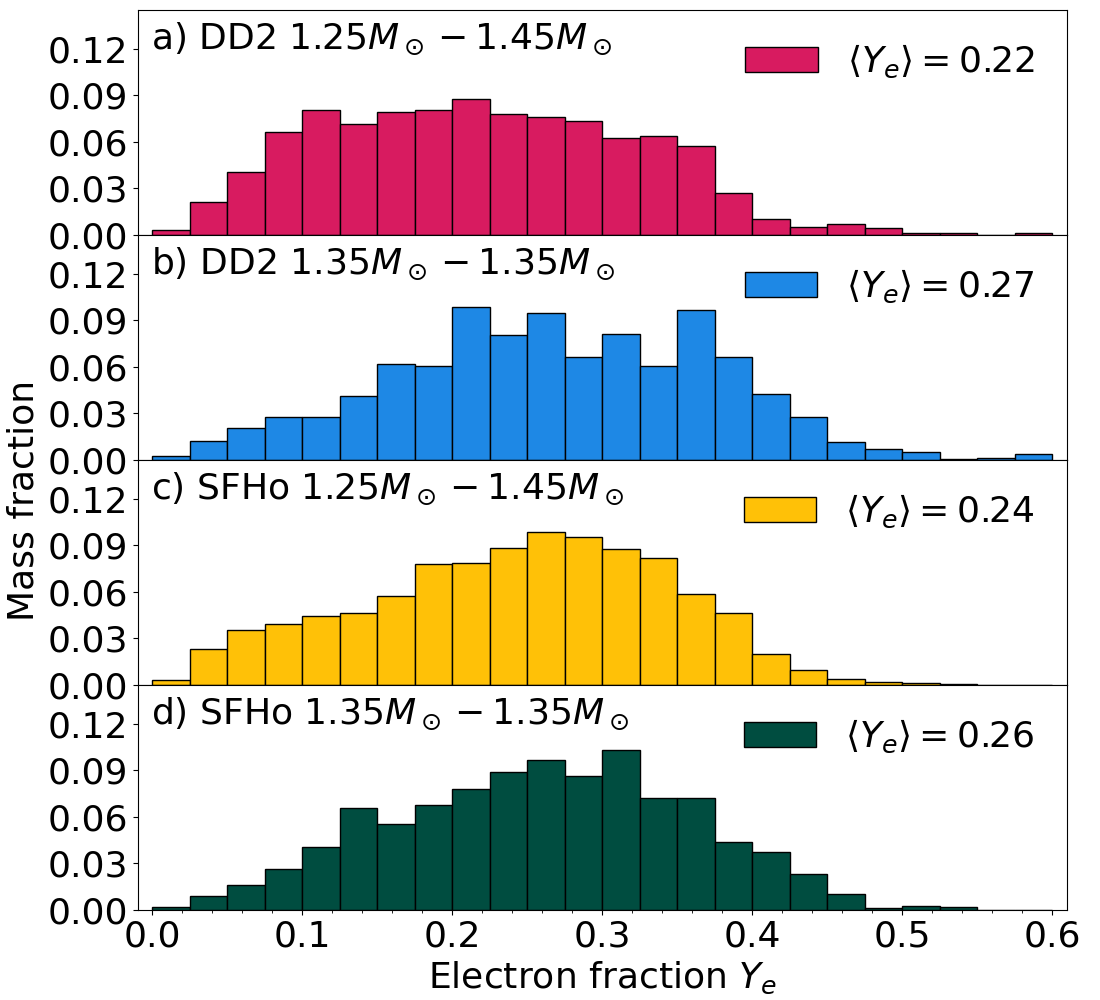}
\caption{(Color online). Fractional mass distributions of the
    matter ejected as a function of $Y_e$ at the time of $\rho=\rho_\mathrm{net}$  together 
    with the mean electron fraction $\langle Y_e \rangle$. From the top: 
    a) DD2-125145, b)  DD2-135135, c) SFHo-125145
    and d) SFHo-135135 NS-NS merger models. 
    }
\label{fig_yedist}
\end{figure}

The corresponding $Y_e$ distributions at $\rho=\rho_\mathrm{net}$, {\it i.e.} initial to the full reaction network computation of the nucleosynthesis, together with the mean value $\langle Y_e \rangle$ are shown in Fig.~\ref{fig_yedist} for the four NS-NS merger models. Each of these distributions are used as initial condition to estimate the final composition of the ejecta through a full reaction network (see Sect.~\ref{sec_nuc}). 

\begin{figure}
\includegraphics[width=\columnwidth]{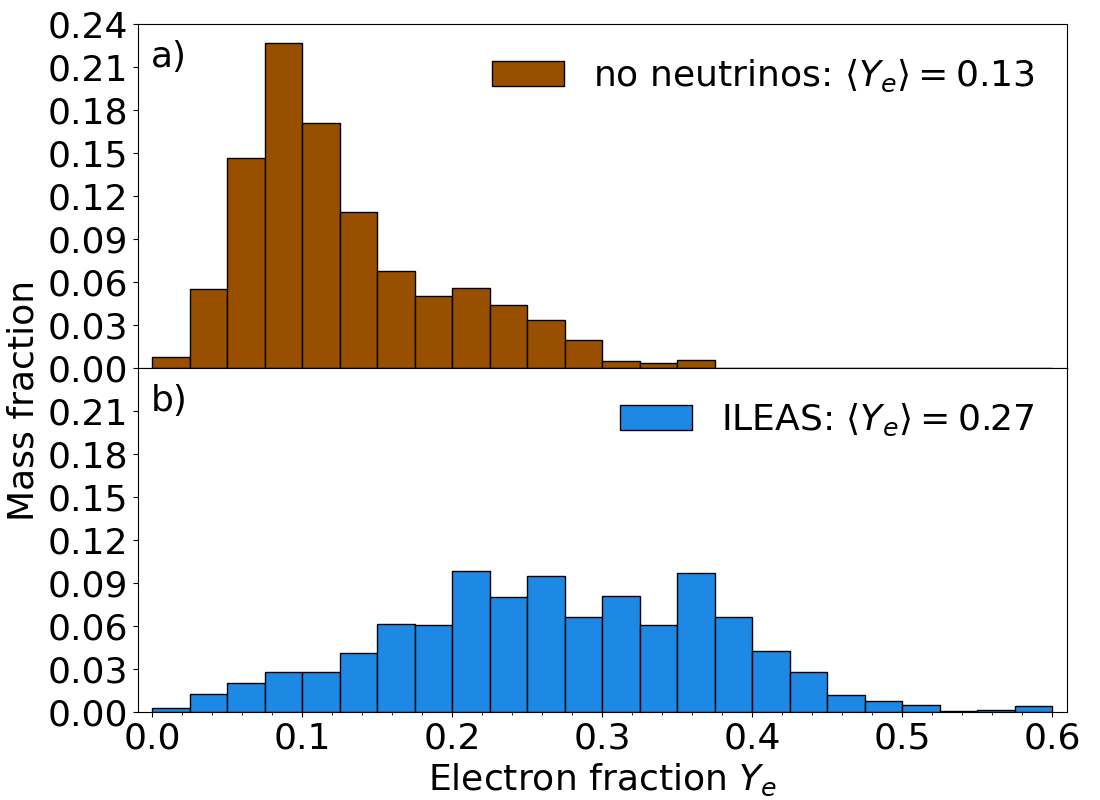}
  \caption{(Color online). Same as Fig.~\ref{fig_yedist} for the two cases, without (a) or with (b) weak nucleonic interactions, of the DD2-135135 model discussed in the text.  Note that the ILEAS case with neutrinos corresponds to Fig.~\ref{fig_yedist}b.
  }
\label{fig_yedist_cases}
\end{figure}

To study the impact of nucleonic weak processes, we have created a reference case in which no such interactions are included after the time when the  lapse function reached a first minimum, which is defined as $t=0$, as in Fig.~\ref{fig_mejec}.
This model follows the same density and temperature evolution as the original DD2-135135 model, but considers the electron fraction at the time  $t=0$, as estimated from the hydrodynamical simulation, and assumes no more weak interaction (Eqs.~\ref{eq:betareac}) with nucleons to take place afterwards. 
Thus, $Y_e$ is assumed to remain unaffected by nucleonic weak processes in all ejected mass elements after $t = 0$. 
This approximation is referred to as the 'no neutrino' case.
In Fig.~\ref{fig_yedist_cases}, the $Y_e$ distributions at $\rho=\rho_\mathrm{net}$ together with the mean values $\langle Y_e \rangle$ are displayed for the DD2-135135 cases with and without neutrinos.  It should be noted that the $Y_e$ distribution without neutrinos does not correspond to the one found in cold NSs  \citep[which was assumed in][]{goriely2011} since matter has been affected by neutrino absorption and emission between the start of the simulation and  time $t=0$. In particular, a significant amount of ejected material has a $Y_e >0.1$, as seen in Fig.~\ref{fig_yedist_cases}a.

Some properties of the ejecta are summarised in Table~\ref{tab:prop}, including in particular the total ejected mass, the mass of its high-velocity component ({\it i.e.} with $v\ge 0.6c$) or the mean electron fraction at $\rho=\rho_{\rm net}$. Additional properties resulting from the nucleosynthesis calculations described in Sect.~\ref{sec_nuc_q} are also given in Table~\ref{tab:prop}. 
These properties are also estimated within the three different angular regions, namely the polar ($0^\circ < |90^\circ - \theta| \leq 30^\circ$), middle ($30^\circ < |90^\circ - \theta| \leq 60^\circ$) or equatorial ($60^\circ < |90^\circ - \theta| \leq 90^\circ$) regions, highlighting the angular dependence of the ejecta composition, as discussed in Sect.~\ref{sec_ang_vc} and of particular relevance to understand the kilonova light curve studied in Part II \citep{just2021b}.

\begin{table*}
\centering
\caption{Summary of the ejecta properties for the five models considered in the present study, including the contribution stemming from three different angular regions, namely the polar ($0^\circ < |90^\circ - \theta| \leq 30^\circ$), middle ($30^\circ < |90^\circ - \theta| \leq 60^\circ$) or equatorial ($60^\circ < |90^\circ - \theta| \leq 90^\circ$)  regions. These properties correspond to the total ejected mass $M_{\rm ej}$, the mass of the high-velocity ejecta $M_{\rm ej}^{v \ge 0.6c}$,  the mean velocity $\langle v/c \rangle$, the mean $Y_e$ at $\rho=\rho_{\rm net}$, the neutron mass fraction $X_n$ at $t=20$~s, the lanthanide plus actinide mass fraction $X_{LA}$ ({\it i.e.} with respect to the total ejecta), the relative amount nuclei $x$ with respect to the mass in each region $x_{A>69}$ and $x_{A>183}$ for r-process nuclei and third-r-process peak nuclei, respectively, and the $^{232}$Th to $^{238}$U ratio. }
\begin{tabular}{llccccccccc}
\hline \hline
Model & Region & $M_{\rm ej}$ & $M_{\rm ej}^{v \ge 0.6c}$ & $\langle v/c \rangle$ & $\langle Y_{e,\rho_{net}}\rangle$ & $X_n^{t=20s}$ & $X_{LA}$ & $x_{A>69}$ & $x_{A>183}$ & Th/U \\
  &   & [$10^{-3}$ \Msun] & [$10^{-4}$ \Msun] &   &   & [$10^{-3}$] &   &   &   &   \\
\hline
  DD2-125145 & Total        & 3.20 & 1.77   & 0.25   & 0.22   & 7.57   & 0.152   & 0.90   & 0.15   & 1.39 \\
             & Equatorial   & 2.20 & 0.91   & 0.24   & 0.19   & 1.60   & 0.126   & 0.97   & 0.30   & 1.40 \\
             & Middle       & 0.78 & 0.62   & 0.24   & 0.26   & 4.22   & 0.022   & 0.95   & 0.11   & 1.35 \\
             & Polar        & 0.22 & 0.23   & 0.27   & 0.33   & 1.76   & 0.004   & 0.78   & 0.05   & 1.30 \\
\hline
  DD2-135135 & Total        & 1.99 & 0.87   & 0.25   & 0.27   & 7.21   & 0.107   & 0.88   & 0.11   & 1.35 \\
             & Equatorial   & 1.35 & 0.46   & 0.25   & 0.25   & 4.71   & 0.086   & 0.95   & 0.17   & 1.36 \\
             & Middle       & 0.45 & 0.23   & 0.24   & 0.29   & 1.72   & 0.017   & 0.93   & 0.09   & 1.30 \\
             & Polar        & 0.20 & 0.18   & 0.27   & 0.34   & 0.78   & 0.003   & 0.75   & 0.07   & 1.33 \\
\hline
 SFHo-125145 & Total        & 8.67 & 2.56   & 0.24   & 0.24   & 2.75   & 0.114   & 0.95   & 0.12   & 1.38 \\
             & Equatorial   & 5.05 & 1.46   & 0.24   & 0.22   & 0.32   & 0.079   & 0.97   & 0.20   & 1.38 \\
             & Middle       & 2.89 & 0.79   & 0.24   & 0.27   & 1.22   & 0.030   & 0.95   & 0.11   & 1.38 \\
             & Polar        & 0.73 & 0.31   & 0.26   & 0.30   & 1.21   & 0.005   & 0.92   & 0.05   & 1.34 \\
\hline
 SFHo-135135 & Total        & 3.31 & 1.53   & 0.29   & 0.26   & 4.76   & 0.115   & 0.92   & 0.11   & 1.35 \\
             & Equatorial   & 2.17 & 0.93   & 0.31   & 0.24   & 0.46   & 0.088   & 0.95   & 0.18   & 1.36 \\
             & Middle       & 0.86 & 0.51   & 0.25   & 0.29   & 2.54   & 0.020   & 0.93   & 0.08   & 1.34 \\
             & Polar        & 0.28 & 0.09   & 0.26   & 0.30   & 1.76   & 0.007   & 0.89   & 0.07   & 1.26 \\
\hline
  DD2-135135 & Total        & 1.99 & 0.87   & 0.25   & 0.13   & 9.17   & 0.246   & 0.97   & 0.43   & 1.31 \\
 no neutrino & Equatorial   & 1.35 & 0.46   & 0.25   & 0.11   & 5.51   & 0.174   & 0.98   & 0.48   & 1.31 \\
             & Middle       & 0.45 & 0.23   & 0.24   & 0.14   & 2.83   & 0.048   & 0.97   & 0.42   & 1.29 \\
             & Polar        & 0.20 & 0.18   & 0.27   & 0.18   & 0.83   & 0.024   & 0.96   & 0.40   & 1.28 \\
\hline \hline\end{tabular}
\label{tab:prop}
\end{table*}

\section{Nucleosynthesis and radioactive decay heat}
\label{sec_nuc_q}

\subsection{Nucleosynthesis}
\label{sec_nuc}

The nucleosynthesis is followed with a full reaction network including all 5000 species from protons up 
to $Z=110$ lying  between the valley of $\beta$-stability and the neutron-drip line \citep[for more details, see][]{goriely2011,bauswein2013,just2015}. All charged-particle 
fusion reactions on light and medium-mass elements that play a role when the nuclear statistical 
equilibrium freezes out are included in addition to radiative neutron captures and photodisintegrations. 
The reaction rates on light species are taken from the NETGEN library, which includes all the latest 
compilations of experimentally determined  reaction rates \citep{xu2013}.  By default,  experimentally unknown reactions 
are estimated  with the TALYS code \citep{goriely2008,koning2012} on the basis of the HFB-21 nuclear masses \citep{goriely2010}, the HFB plus combinatorial nuclear level densities \citep{goriely2008} and the QRPA E1 strength functions \citep{goriely2004}. 
Fission and $\beta$-decay processes, including neutron-induced fission, spontaneous fission, $\beta$-delayed 
fission, as well as $\beta$-delayed neutron emission, are considered as detailed in \citet{goriely2015}.  All fission processes are estimated on the basis of the HFB-14 
fission paths \citep{goriely2007} and the full calculation of the corresponding barrier penetration 
\citep{goriely2009}.  The fission fragment distribution is taken from the microscopic scission-point model,  known as the SPY model,  as described 
in \citet{lemaitre2019}.  The  $\beta$-decay processes are taken from the mean field plus relativistic QRPA calculation of \citet{marketin2016},  when not available 
experimentally. This nuclear physics set represents our standard input. A sensitivity analysis of our results to the nuclear ingredients is postponed to a future study.

\begin{figure}
\includegraphics[width=\columnwidth]{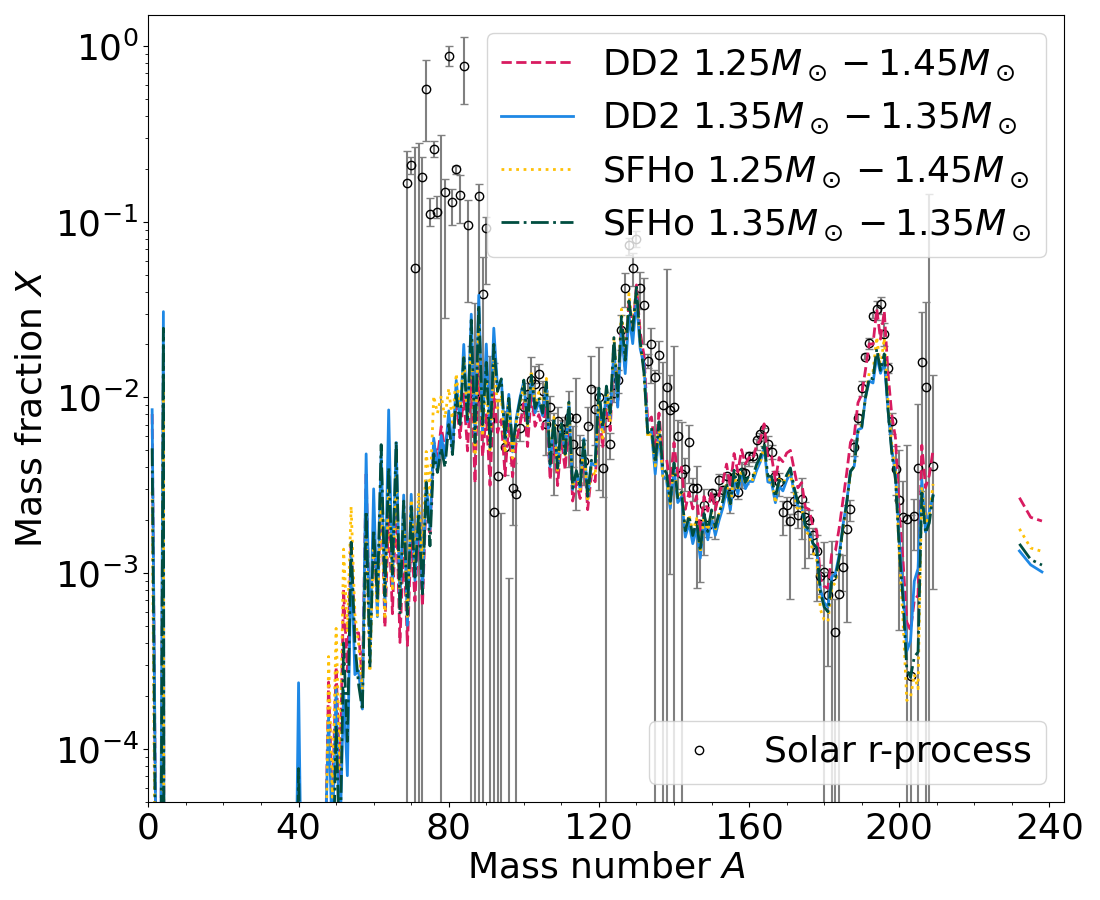}
  \caption{(Color online).  Final mass fractions of the material ejected as a function of the atomic mass $A$ for our DD2 1.25--1.45\Msun\,  DD2 1.35--1.35\Msun\, SFHo 1.25--1.45\Msun\ and  SFHo 1.35--1.35\Msun\ NS-NS merger models.  The solar system r-abundance distribution  (open circles) from \citet{goriely1999} is shown for comparison and arbitrarily normalised to the DD2 asymmetric model at the third r-process peak ($A\simeq 195$).}
\label{fig_rpro_aa}
\end{figure}

Fig.~\ref{fig_rpro_aa} shows the final isotopic abundance distributions obtained if we adopt the initial ILEAS $Y_e$ distributions of Fig.~\ref{fig_yedist} for the four hydrodynamical merger models. Given the similar and relatively wide initial $Y_e$ distributions, the resulting abundance distributions are almost identical and reproduce rather well the solar system r-abundance distribution above $A\ga 90$. For all models, we have an efficient r-process nucleosynthesis with the production of lanthanides, second- and third-peak nuclei. The lanthanide plus actinide mass fraction $X_{LA}$, the relative amount of r-process nuclei $x_{A>69}$ and of third-r-process peak nuclei $x_{A>183}$,({\it i.e.} with $A>183$) are summarised in Table~\ref{tab:prop} for each model. In particular, the ejecta of all four systems can be seen to consist of 88 up to 95\% of $A>69$ r-process material with lanthanides plus actinides ranging between 11 and 15\% in mass. In all four models, the third r-process peak is rather well produced and includes between 11 to 15\% of the total mass. The DD2-125145 model has a relatively larger production of the heaviest r-process elements, as indicated by a larger value of $x_{A>183}$, which reaches about 30\% in the equatorial region.

\begin{figure}
\includegraphics[width=\columnwidth]{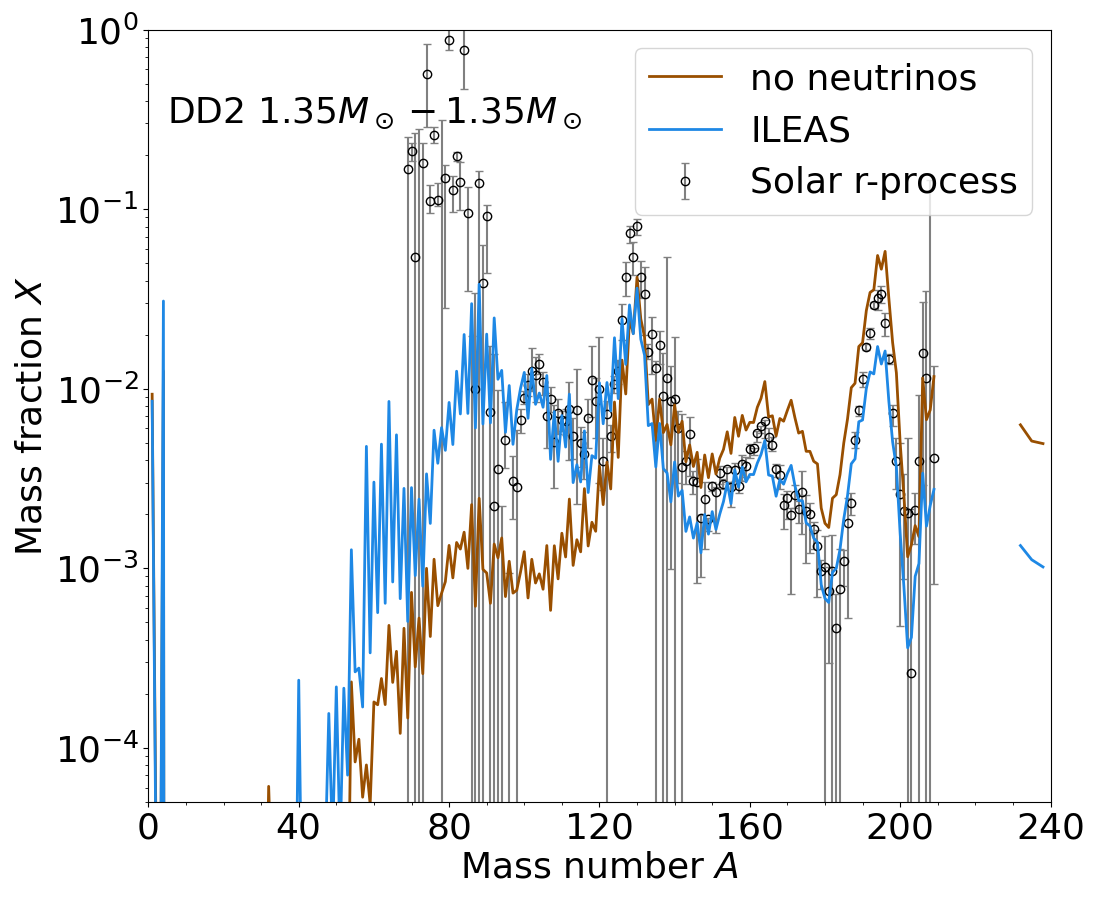}
  \caption{(Color online).  Mass fractions of the $2\times 10^{-3}$\Msun\ of material
    ejected in our 1.35--1.35\Msun\ NS-NS merger model with DD2 EoS as a function of the atomic mass $A$ for the two cases studied here, {\it i.e} with (ILEAS) or without (no neutrinos) weak nucleonic interactions.  The solar system abundance distribution is normalised as in Fig.~\ref{fig_rpro_aa}. }
\label{fig_rpro_aa_cases}
\end{figure}

For the DD2-135135 model, Fig.~\ref{fig_rpro_aa_cases} shows the final isotopic abundance distributions of the case without neutrinos compared to the case where neutrino interactions are included.
 If we assume the initial $Y_e$ distribution to be unaffected by weak interactions (Fig.~\ref{fig_yedist_cases}a),  the resulting distribution is characteristic of what has been obtained by most of the calculations neglecting neutrino absorption, {\it i.e.} the production of $A \ga 130-140$ is considerably enhanced due to the dominance of $Y_e<0.1$ trajectories and an efficient fission recycling. The production of $A\simeq 130$ nuclei in the second r-process peak is linked to the non-negligible presence of $Y_e> 0.15 $ trajectories (see Fig.~\ref{fig_yedist_cases} and the discussion in Sect.~\ref{sect_weak}). The ``no neutrino'' case is found to be composed of 2.1 (4.7) times more lanthanides (actinides) and a significantly more pronounced third r-process peak (Table ~\ref{tab:prop}).

\begin{figure}
\includegraphics[width=\columnwidth]{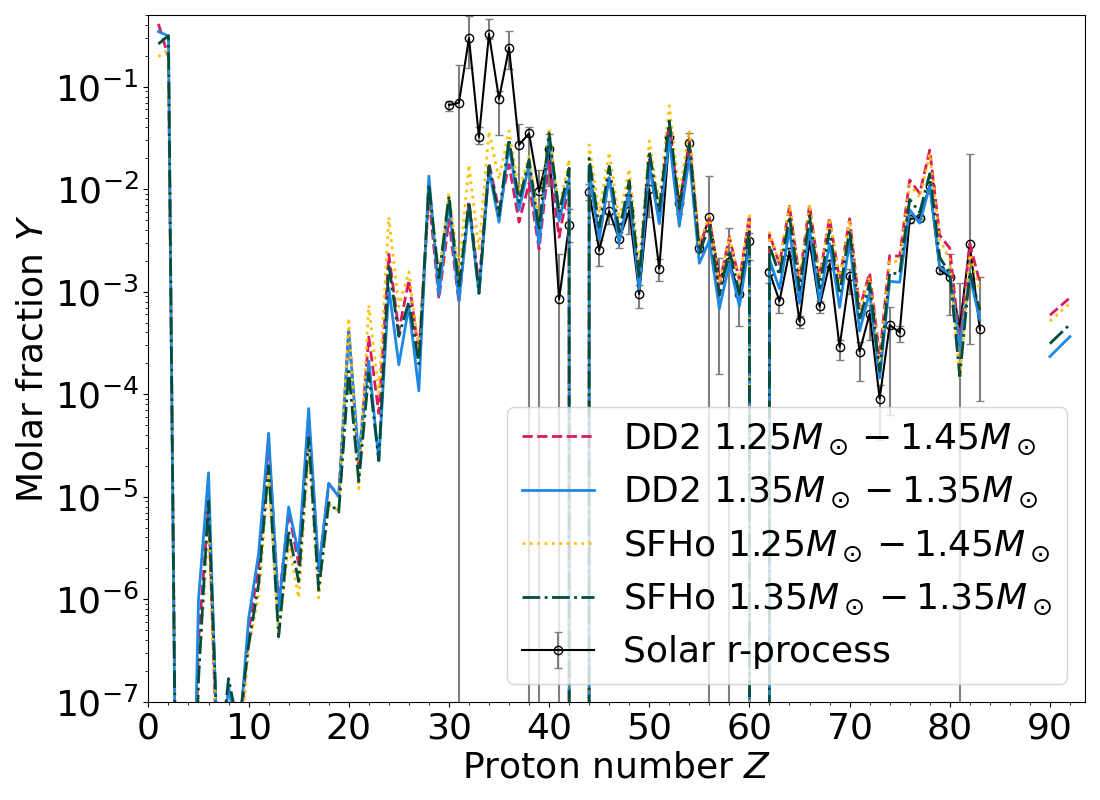}
  \caption{(Color online). Same as Fig.~\ref{fig_rpro_aa} for the elemental abundance distributions given by the molar fractions. The solar distribution is normalized to the DD2-125145 prediction of the third r-process peak.}
\label{fig_rpro_zz}
\end{figure}

The final elemental abundance distributions obtained from the four hydrodynamical models including weak processes are shown in Fig.~\ref{fig_rpro_zz}. As for the isotopic distributions, there are only minor differences between the four elemental distributions. In particular,  the production of actinides is larger for the two asymmetric merger models. However, the $^{232}$Th to $^{238}$U ratio remains rather constant and equal to 1.35--1.39 for all four models (Table~\ref{tab:prop}), a property of particular interest to cosmochronometry \citep[e.g][]{goriely2016}.

\begin{figure}
\includegraphics[width=\columnwidth]{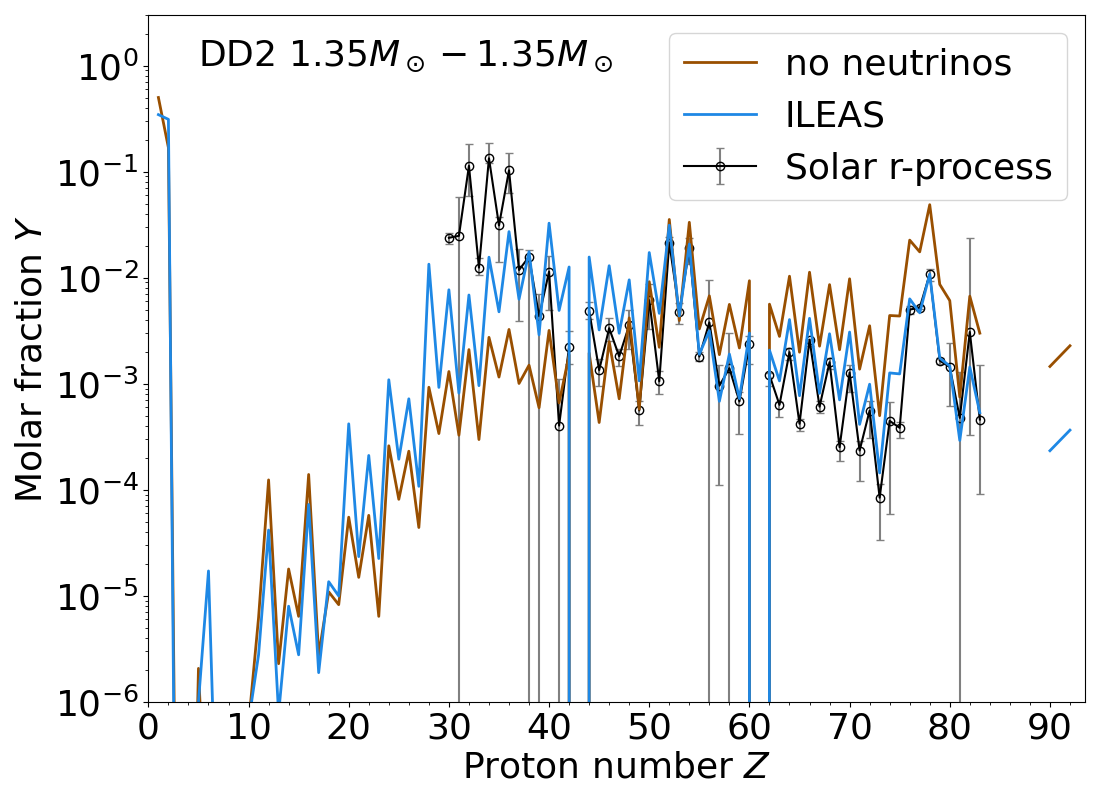}
  \caption{(Color online). Same as Fig.~\ref{fig_rpro_aa_cases} for the elemental abundance distributions given by the molar fractions. The solar distribution is normalized as in Fig.~\ref{fig_rpro_zz}.}
\label{fig_rpro_zz_cases}
\end{figure}

The elemental distributions of the DD2-135135 cases with and without neutrinos are presented in Fig.~\ref{fig_rpro_zz_cases}.  We can see that a rather different prediction is obtained when including weak processes, in particular, a significantly smaller amount of $Z \ga 50$ elements is produced.  
However, although the actinide production for the ILEAS case is significantly smaller compared to the reference neutrino-less simulation, the elemental ratio Th/U remains rather constant.  
For the ILEAS case, the Ni to Zr region dominates the ejecta,  and the production of Sr ($Z=38$) is 14 times larger compared to the case without neutrinos.  Sr is of special interest after its identification by \citet{watson2019} in the AT2017gfo spectrum.  Such a different elemental distribution impacts the observed kilonova light curve,  as discussed in Part II \citep{just2021b}.


\subsection{Radioactive decay heat}
\label{sec_q}
\begin{figure}
\includegraphics[width=\columnwidth]{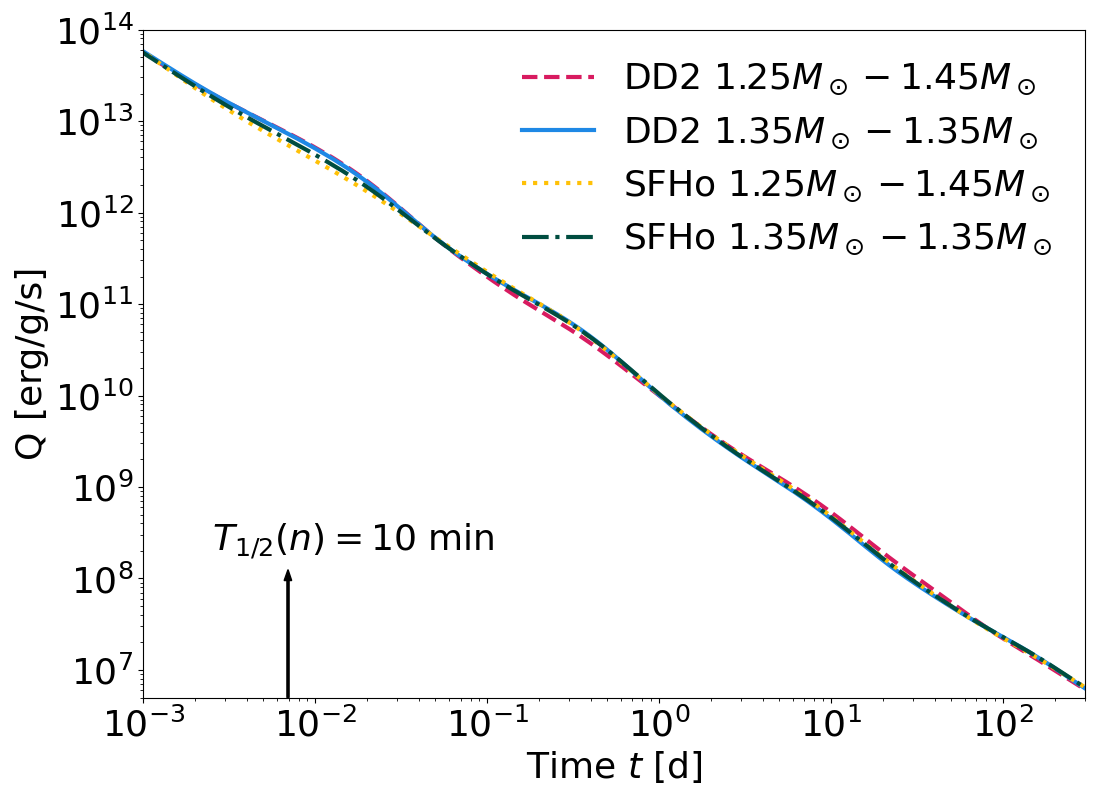}
  \caption{(Color online). Time evolution of the radioactive heating rate $Q$  for the four hydrodynamical models including weak interactions. The arrow at $t=10$~min corresponds to the decay half-life of the neutron.}
\label{fig_Q_t}
\end{figure}

The energy release by radioactive decay is estimated consistently within the same nucleosynthesis network for each of the models considered. No thermalisation efficiency is included at this stage and only the contribution from neutrino energy loss in each of the $\beta$-decays are removed following the prescription of \citet{Fowler1975},  
{\it i.e.} what we denote as radioactive heating rate $Q$ in this study is the total energy carried by $\alpha$, $\beta$, and $\gamma$ particles as well as fission products.  
Fig.~\ref{fig_Q_t} shows the time evolution of the radioactive heating rate $Q$ for the four hydrodynamical models including weak interactions. 
As indicated by the arrow, a bump corresponding to the ejection of free neutrons that decay after $t=10$~min is visible. 
In particular, the models using the DD2 EoS have the largest ejection of free neutrons with a mass fraction of 0.7\% at late time $t\simeq 20$~s (see  Table~\ref{tab:prop}). 
The absence of neutrino interactions would provide even larger quantities, typically a $\sim 30$\% increase of ejected free neutrons. 
For both SFHo models the ejected amount of free neutrons is smaller, corresponding to 0.3\% and 0.5\% of the ejected material for the SFHo-125145 and SFHo-135135 models, respectively. 
In these cases, lower ejection velocities in comparison to the DD2 models are found, and therefore a less noticeable feature linked to neutron decay is observed.
The decay of free neutrons may power a precursor signal of the kilonova,  as discussed in \citet{metzger2015}. The inclusion of neutrino interaction may, however, decrease its strength, though it is essentially linked to the very fast ejection of a few neutron-rich mass elements.  

\begin{figure}
\includegraphics[width=\columnwidth]{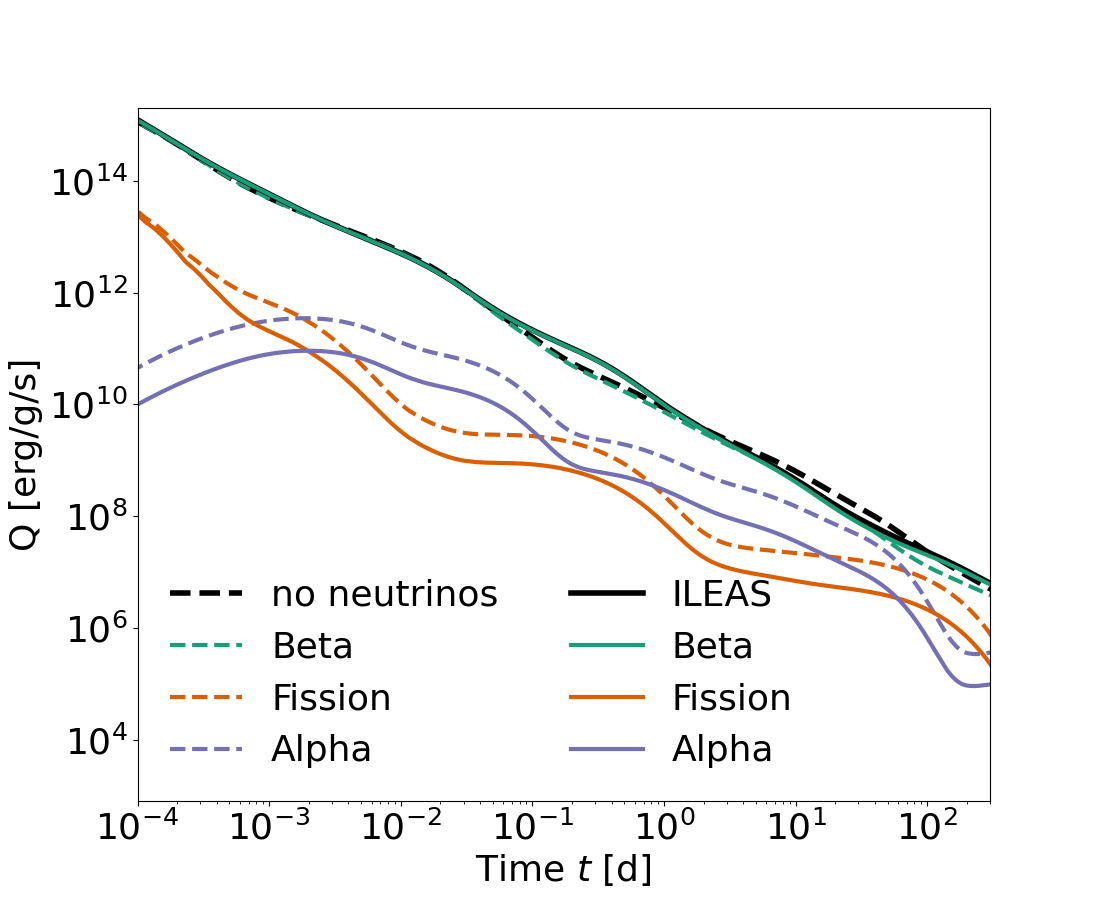}
  \caption{(Color online). Same as Fig.~\ref{fig_Q_t} for the symmetric DD2 model with (solid lines) and without (dashed lines) weak nucleonic interactions. The total heating rate is shown in black, the $\beta$-, $\alpha$- and fission contributions in green, orange and purple colours, respectively. }
\label{fig_Q_t_cases}
\end{figure}

In general, the heating rate is dominated by a large number of neutron-rich $A \la 200$ nuclei that $\beta$-decay towards the valley of stability. Even after several days, the main contribution to the heating rate comes from a few $\beta$-unstable isotopes with half-lives on the order of days. However, at late time ($t>10$~d), the contribution from $\alpha$-decay and spontaneous fission of trans-Pb species starts to become significant (Fig.~\ref{fig_Q_t_cases}). 
More specifically, the $\alpha$-decay chains starting from $^{222}$Rn, $^{223-224}$Ra and, in particular, $^{225}$Ac significantly contribute to the heating rate at time $t>10$~d.  In addition,  the $\alpha$-decay of $^{253}$Es and $^{255}$Es produces an important amount of heat, however,  several orders of magnitude less than the four aforementioned-mentioned $\alpha$-decay chains since their daughter nuclei quickly $\beta$-decay to long-lived Cf. 
Among the fissioning species, $^{254}$Cf is the only notable isotope, in fact, it has one of the largest heating rates across all decay modes and isotopes for $t>10$~d. 
The production of trans-Pb are often taken as  the signature of r-processing beyond the heaviest stable elements \citep{wanajo2018,wu2019,zhu2018}.
Among our four hydrodynamical models, it is the asymmetric DD2 model that shows the most pronounced contribution to the heating rate by trans-Pb species for  $t > 10$~d, as also shown by the large production of Th and U in Fig.~\ref{fig_rpro_aa}. 

\begin{figure}
\includegraphics[width=\columnwidth]{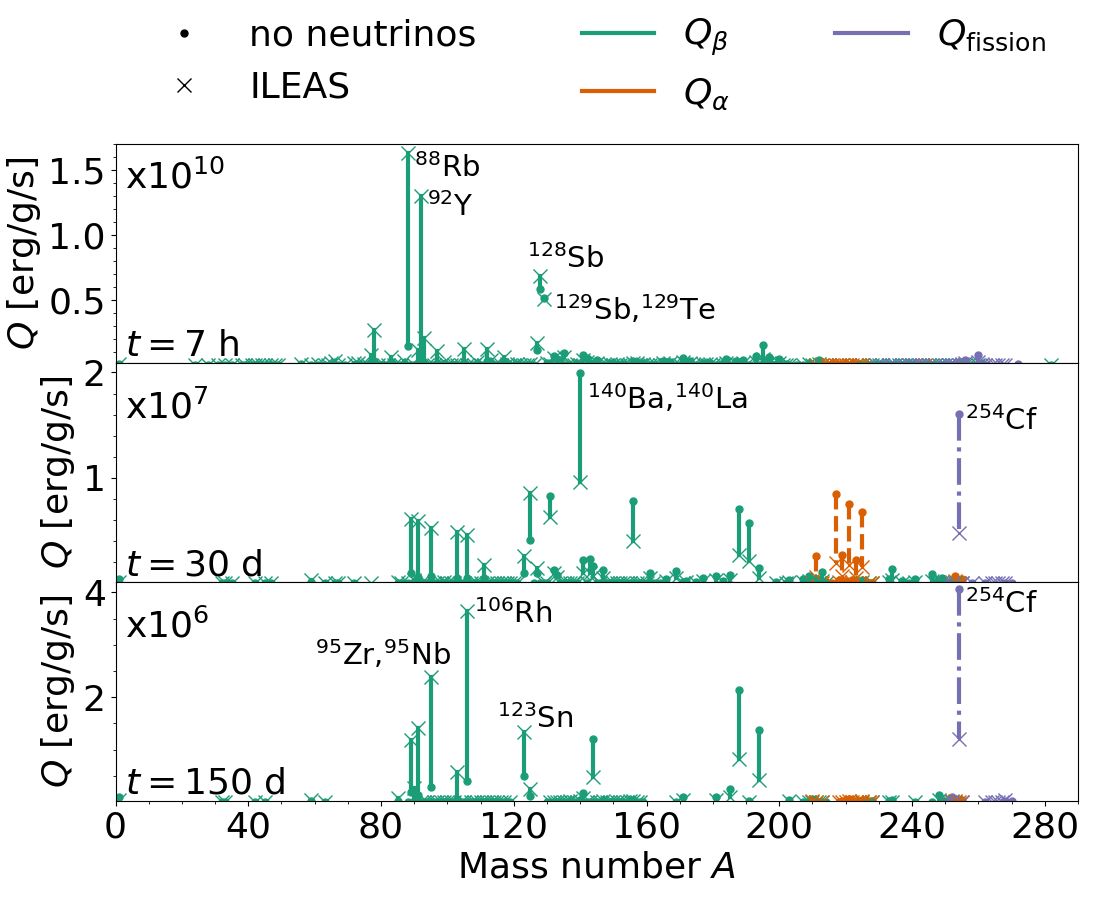}
  \caption{(Color online).  Contribution to the radioactive heating rate $Q$ stemming from individual nuclear decays versus mass number $A$ at $t=7$~h, 30~d and 150~d in the top, middle and bottom panels, respectively. The no-neutrinos model is represented by a dot and the ILEAS model by a cross and both predictions are connected by a line. The colours represent the decay modes as in Fig.~\ref{fig_Q_t_cases}.  }
\label{fig_dQ}
\end{figure}

In Fig.~\ref{fig_Q_t_cases} the heating rate of our two models with and without neutrino interactions are compared, highlighting the relative contribution from the main decay modes $\beta$, $\alpha$ and fission.  In general, the time evolution of the heating rate due to $\alpha$-decay and spontaneous fission follow the same trend for both models, where the heating rate of the no-neutrinos model is up to one order of magnitude higher. For the total heating rate, the main differences occur around $t\sim7$~h, 30~d and after 150~d (see Fig.~\ref{fig_dQ} for the heating rate versus mass number at these three time points). 
At 7~h after the merger,  the discrepancies are caused by an excess of heat for the ILEAS model mainly generated by the $\beta$-decay of  $^{88}$Kr-$^{88}$Rb and $^{92}$Y (top panel Fig.~\ref{fig_dQ}).  
This can be seen in Fig.~\ref{fig_rpro_zz_cases} where an excess amount of elements in the $33 \le Z \le\ 39$ region are produced in the simulation including neutrinos and are responsible for the large radioactive heat through $\beta$-decays at times between 0.1 and 1 day (Fig.~\ref{fig_Q_t_cases}).  The same effect can be seen after $t>60$~d where in the no-neutrinos case $\beta$-decay heating rate starts to decrease, while this does not occur within the ILEAS model due to heat generated by $A<130$ nuclei with longer half-lives. 
After $t>10$d the signature of trans-actinides production start to differ significantly. The ILEAS model is still dominated by $\beta$-decays of $A \la 200$ species, however,  for the no-neutrinos model the contribution from the $A=222-225$ $\alpha$-decay chains and spontaneous fission of $^{254}$Cf becomes important.  In addition, the case without weak interactions has a larger production of elements with $A>130$ (see Fig.~\ref{fig_rpro_zz_cases}), leading to more heat generated by $\beta$-decays, in particular by $^{140}$Ba and $^{140}$La  (see middle panel Fig.~\ref{fig_dQ}).
Later at $t=150$~d (bottom panel Fig.~\ref{fig_dQ}), the major trans-actinide contributor is $^{254}$Cf, with a heating rate about 4 times larger in the no-neutrinos model, while the $\beta$-decay of $A<130$ nuclides powers the heating rate in the ILEAS model.

\subsection{Sensitivity to the electron fraction}
\label{sect_sens}

\begin{figure}
\includegraphics[width=\columnwidth]{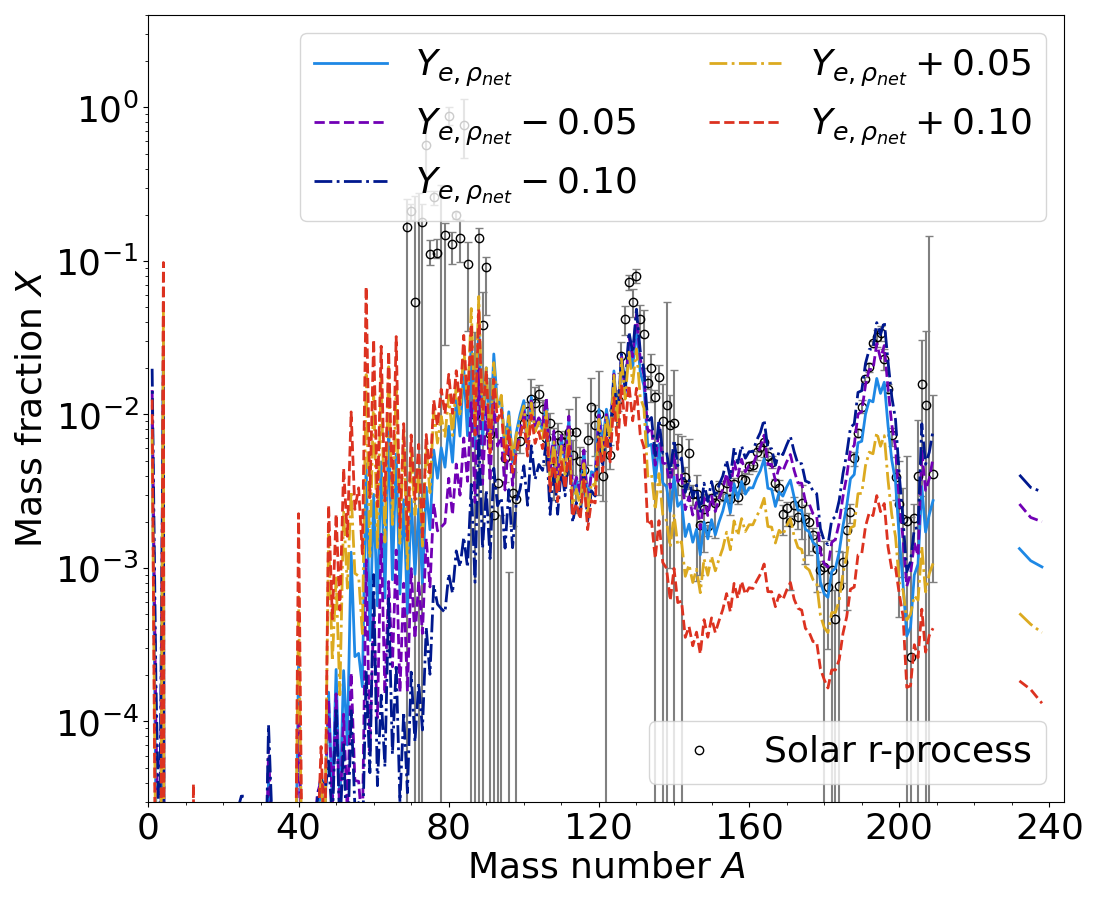}
  \caption{(Color online). Final abundance distribution in the dynamical ejecta of the DD2-135135 model assuming the initial $Y_e$ is systematically modified with respect to ILEAS prediction by $\pm 0.05$ or $\pm 0.10$. The solar system abundance distribution is normalised as in Fig.~\ref{fig_rpro_aa}.}
\label{fig_rpro_dye}
\end{figure}

\begin{figure}
\includegraphics[width=\columnwidth]{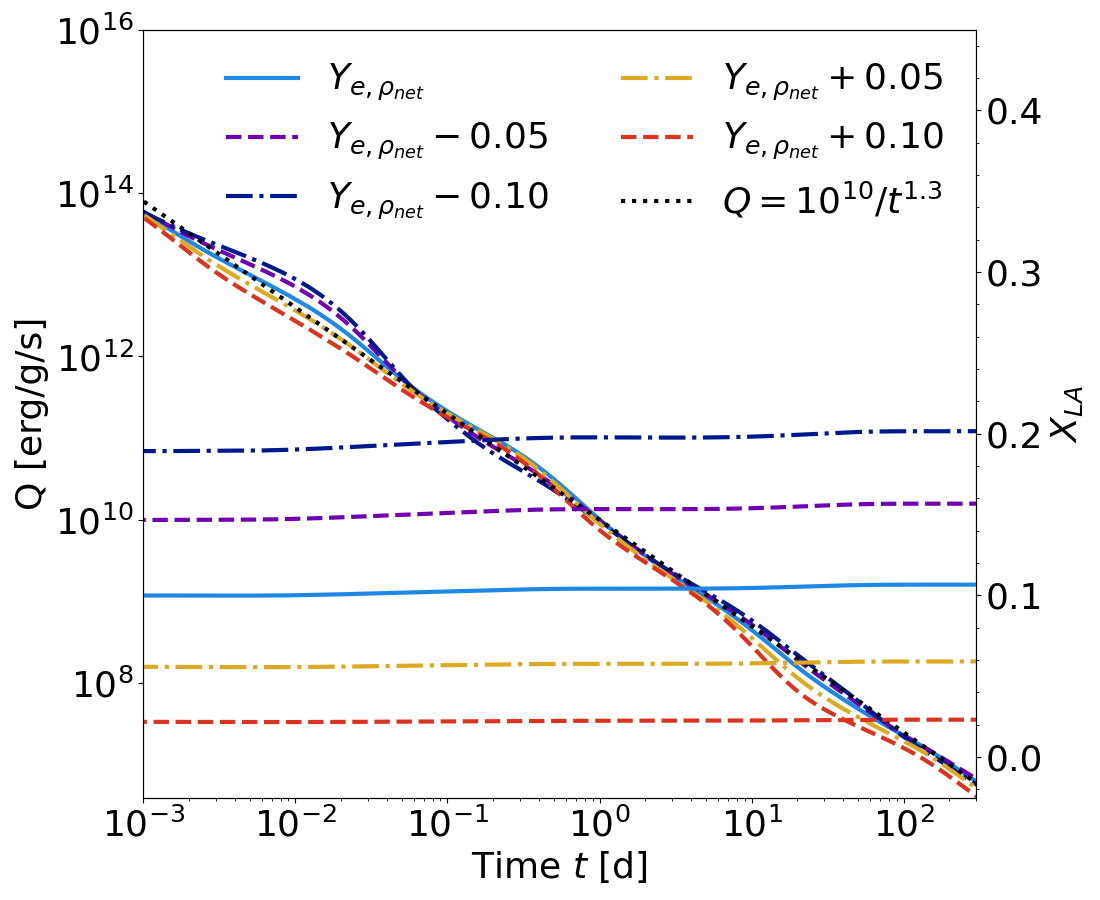}
  \caption{(Color online). Time evolution of the radioactive heating rate $Q$  and the lanthanide+actinide mass fraction $X_{LA}$ for the DD2-135135 model assuming the initial $Y_e$ is systematically modified with respect to ILEAS prediction by $\pm 0.05$ or $\pm 0.10$. The black dotted line corresponds to the approximation $Q_0=10^{10}~(t[{\rm d}])^{-1.3}$ \citep{metzger2010}.}
\label{fig_Q_dye}
\end{figure}

As briefly discussed in Sect.~\ref{sect_weak}, NS merger simulations still suffer from numerous uncertainties connected to, among others, the neutrino transport and the equation of state. Those uncertainties inevitably affect the initial conditions for the nucleosynthesis calculations within the model described in Sect.~\ref{sect_weak}. In particular, the initial $Y_e$ at the time of $\rho=\rho_\mathrm{net}$ may differ by $\pm 0.05$ or even $\pm 0.10$ with respect to the present modelling. To test the sensitivity with respect to the many uncertainties affecting still the hydrodynamical simulations, we have considered extreme cases in which the  $Y_e$ at the time of $\rho=\rho_\mathrm{net}$ is systematically increased or decreased by 0.05 or even 0.10. We show in Figs.~\ref{fig_rpro_dye}-\ref{fig_Q_dye} the impact of such different initial conditions on the final composition of the ejected material as well as the time evolution of the radioactive heating rate and the lanthanide+actinide mass fraction $X_{LA}$ for the DD2-135135 model.

As seen in Fig.~\ref{fig_rpro_dye}, a systematic increase of $Y_e$ by 0.1 is not sufficient to give rise to the production of the first r-process peak, but rather favours the synthesis of the Fe, Ni or Zn  even-$N$ isotopes in the $A=58-66$ region, as well as $^{88}$Sr and $^{84,86}$Kr. As a consequence, a significant reduction in the production of the second and third r-process peaks, as well as lanthanides and actinides, can be observed (Fig.~\ref{fig_Q_dye}). 
Note, however, that even in the least neutron-rich case where $Y_e$ is increased by 0.10 the mass fraction of $X_{LA}\simeq 0.02$ remains not negligible.
The radioactive decay heat $Q$ is found to follow rather well the empirical approximation $Q_0=10^{10}~t[{\rm d}]^{-1.3}$ \citep{metzger2010} in all the 5 cases shown in Fig.~\ref{fig_Q_dye}.  The differences seen in the various $Y_e$ cases in Fig.~\ref{fig_Q_dye} can be explained by the same effects as those found in the  comparison between the cases with and without weak interactions described in Sect.~ \ref{sec_q}.
In the most neutron-rich case ($Y_e-0.10$), the $Q$ enhancement around $t=10^{-2}$~d and $t>10$~d is due to the decay of free neutrons and actinides, respectively. 
For the least neutron-rich case, the trans-actinides signal at $t>10$~d completely disappears and the heating rate is essentially powered by $\beta$-decay of $A\leq 140$ nuclei, except for the spontaneous fission of $^{254}$Cf. 

The highest $Y_e+0.10$ case with the lowest $X_{LA}$ gives rise to a lower heating rate $Q$ with respect to the most neutron-rich case, in contrast to what is obtained by \citet{martin2015} (see their Fig.13) who found,  relative to $Q_0$, a heating rate enhancement by a factor of 2.5 at $t\sim 4$~h for their neutrino-driven wind producing the light r-process elements. Our heating rate is also significantly lower than the value found by \citet{perego2017} (see in particular their Eq. 2, and Part II for a detailed comparison with their heating rates) for $Y_e \ga 0.25$ ejecta. 

\section{Angular and velocity dependence of the ejecta}
\label{sec_ang_vc}

\begin{figure*}
\begin{center}
\includegraphics[width=0.99\textwidth]{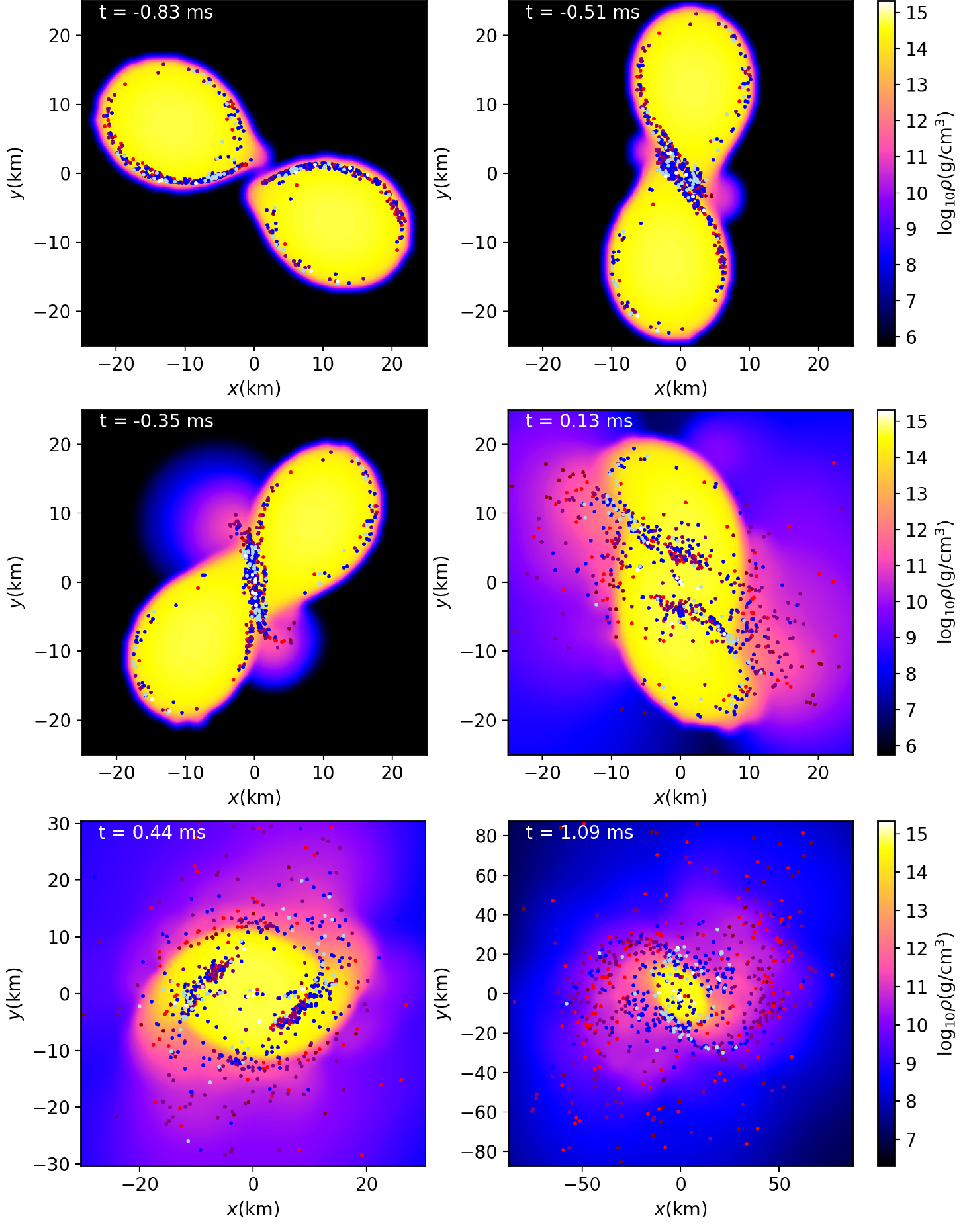}
\caption{(Color online).  2D snapshots of the density distributions of the DD2-135135 simulation in the equatorial plane at six different times.
The locations of the 783 finally ejected mass elements are indicated by coloured dots, projected onto the equatorial plane, with the final $Y_e$-values colour-coded within the following ranges: dark
red: $Y_e < 0.1$, red: $0.1 \le Y_e < 0.2$, purple: $0.2 \le Y_e < 0.3$, blue: $0.3 \le Y_e < 0.4$, light blue: $0.4 \le Y_e < 0.5$, white: $Y_e \ge 0.5$. Note that the high-$Y_e$ particles move mostly perpendicular to the plane of the figure.
}
\label{fig_snapshot}
\end{center}
\end{figure*}

\begin{figure}
\includegraphics[width=\columnwidth]{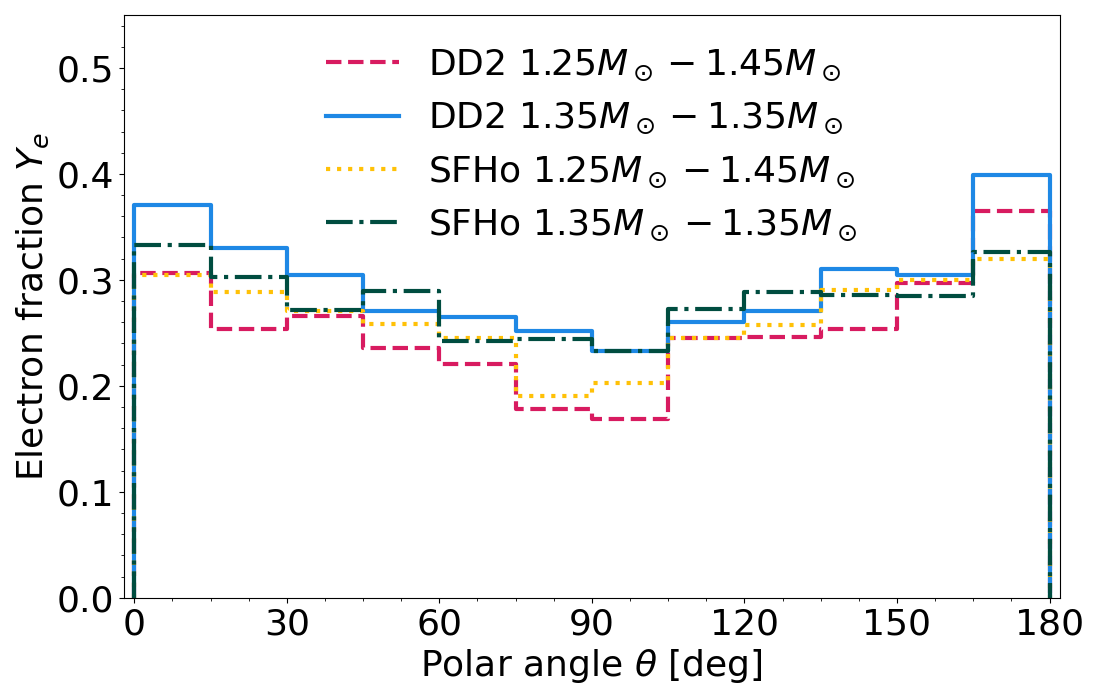}
  \caption{(Color online).  Average electron fraction per polar angle bin $d\theta$
   at $\rho=\rho_\mathrm{net}$ 
  as predicted by the ILEAS hydrodynamical simulation for the four NS-NS merger models: 
  DD2-125145, DD2-135135, SFHo-125145 and SFHo-135135. 
  }
\label{fig_yedist_ang}
\end{figure}

The three-dimensional implementation of the weak nucleonic reactions (Eqs. \ref{eq:betareac}) within the ILEAS framework of the hydrodynamical simulations allows us to study the angle and velocity distributions of the ejecta and nucleosynthesis yields. 
As shown in Fig.~\ref{fig_snapshot} and by previous studies \citep[see e.g.][]{goriely2011,sekiguchi2015}, two major mechanisms during the merger phase are responsible for the dynamical mass ejection, namely tidal stripping preferentially in the orbital plane and the more isotropic mass ejection by shock compression at the NS contact interface.
Most of the high-$Y_e$ mass elements are seen to originate from the collision interface. 
In the equatorial plane mostly lower-$Y_e$ material is ejected, whereas regions at higher polar angles contain a mix of low-$Y_e$ and high-$Y_e$ ejecta.
Indeed, Fig.~\ref{fig_yedist_ang} displays the electron fraction distribution of the ejected matter as a function of polar angle $\theta$ for the four merger models. It can be seen that the material at the poles is significantly less neutron-rich, in particular for the DD2-125145, SFHo-125145 and DD2-135135 models. It is consequently of particular interest for observation \citep[see][]{just2021b} to study in more detail the composition and decay heating rate obtained as a function of the polar angle, but also as a function of the expansion velocity. Some properties of the ejecta in the different angular regions, namely the polar, middle or equatorial regions, are given in Table~\ref{tab:prop}.
In the following section a detailed analysis of the angle and velocity distribution of the material as well as the resulting nucleosynthesis is presented for the DD2-135135 model. If the other models deviate from its trend, a special mention will be made.

\begin{figure}
\includegraphics[width=\columnwidth]{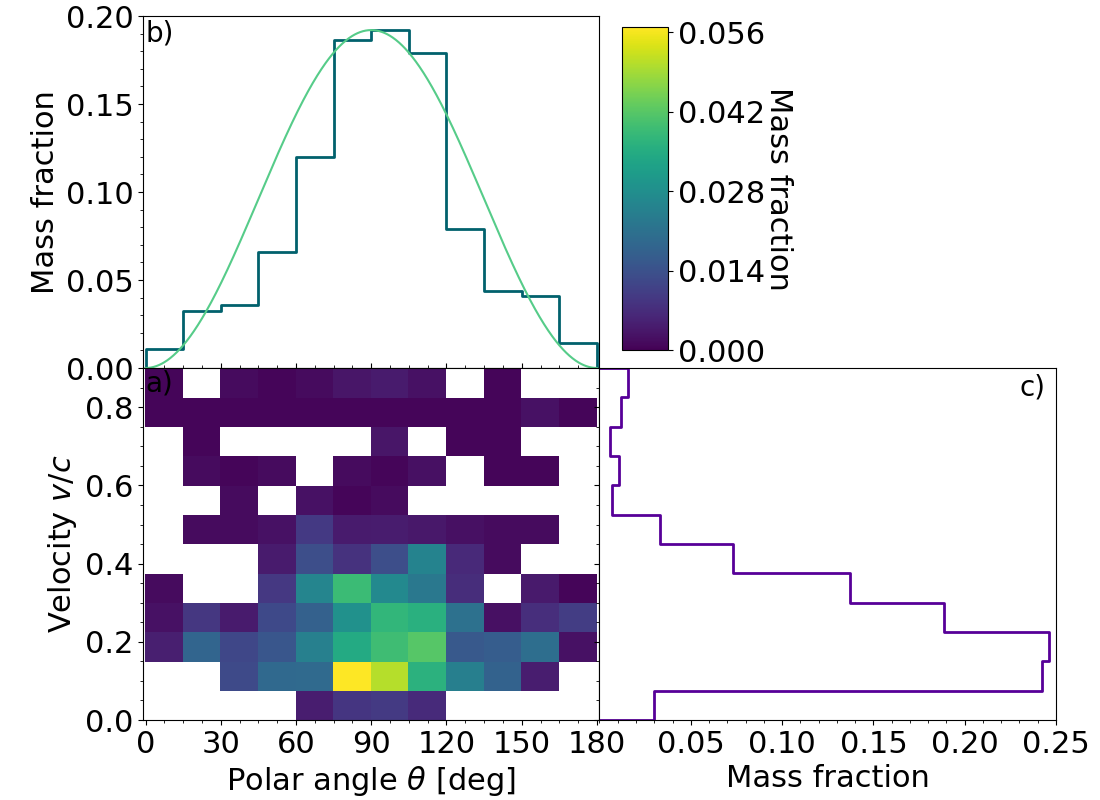}
  \caption{(Color online).  Mass distribution of the ejecta a) in the two-dimensional plane of the polar 
  angle v.s. velocity at the end of the DD2-135135 NS-NS hydrodynamical simulation. b) and c) panels represent the one-dimensional projections of the integrated ejected masses on the $\theta$ and $v/c$ axes, respectively. In panel a), the white
regions depict the absence of any  trajectory. The green line in panel b) corresponds to the $\rm{sin}^2(\theta)$ function, arbitrarily normalized.
  }
\label{fig_dM2d}
\end{figure}

Fig.~\ref{fig_dM2d}a illustrates the mass distribution of the ejecta in the two-dimensional plane of the polar angle $\theta$ and  velocity $v/c$.  Figs.~\ref{fig_dM2d}b and \ref{fig_dM2d}c give the corresponding projection of the integrated mass fraction on the $\theta$ and $v/c$ axes, respectively.  Fig.~\ref{fig_dM2d}b shows that most of the material is ejected in the equatorial region ($60^\circ < \theta \leq 120^\circ$) of the NSs merger plane. The angular distribution is seen to be roughly following a $\rm{sin}^2(\theta)$ function, as pointed out by \citet{perego2017}. However, compared to the $\rm{sin}^2(\theta)$ distribution the simulations tend to produce a pattern that is slightly more peaked towards $\theta=90^\circ$.

In Fig.~\ref{fig_dM2d}c, we see that most of the ejected material has a velocity below $0.3c$, and that the high-velocity tail ($v/c>0.6$) contains only a minor fraction of the total ejected mass (between 1-5\% of the total mass, see also Table~\ref{tab:prop}) and is represented by 63 (22) and 37 (13) particles for the DD2 and SFHo asymmetric (symmetric) models, respectively. These fast-moving particles contain over 99\% of the $X_n^{t=20\mathrm{s}}$ mass.
The high-velocity ejecta are not isotropic, but more concentrated in the equatorial plane, though they are distributed over all angles, especially for the DD2 EoS models (Fig.~\ref{fig_dM2d}a).

\begin{figure}
\includegraphics[width=\columnwidth]{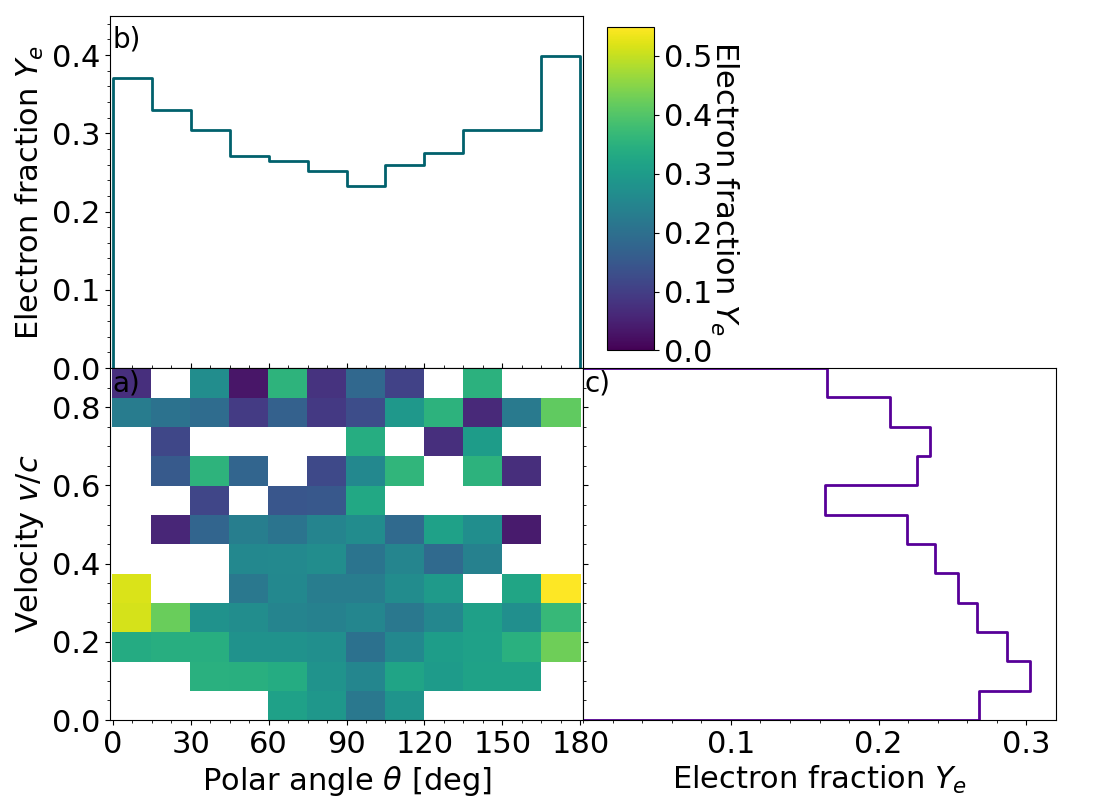}
  \caption{(Color online).  Same as Fig.~\ref{fig_dM2d} for the electron fraction distribution at $\rho=\rho_\mathrm{net}$ 
  as predicted by the ILEAS hydrodynamical simulation. b) and c) panels represent the mass-averaged $Y_e$ along the $\theta$ and $v/c$ axes, respectively.
  }
\label{fig_ye2d}
\end{figure}

In general, the further away from the equatorial plane the mass element, the larger its minimum velocity. In particular,  the minimum velocity in the extreme polar regions ($\theta \simeq 0^\circ$ and $\theta \simeq 180^\circ$) is larger than $0.2c$, while $v/c\simeq 0.03$ is the lower limit in the equatorial plane.
Fig.~\ref{fig_ye2d}a shows the electron fraction distribution in the $\theta-v/c$ plane, and Figs.~\ref{fig_ye2d}b and \ref{fig_ye2d}c the projection of the average $Y_e$ on the $\theta$ and $v/c$ axes, respectively.  The ejecta in the equatorial regions is the most neutron rich.  As seen in Fig.~\ref{fig_ye2d}c, most of the high-$Y_e$ mass elements have velocities in the range $\sim 0.1-0.4c$.  

%

\begin{figure}
\includegraphics[width=\columnwidth]{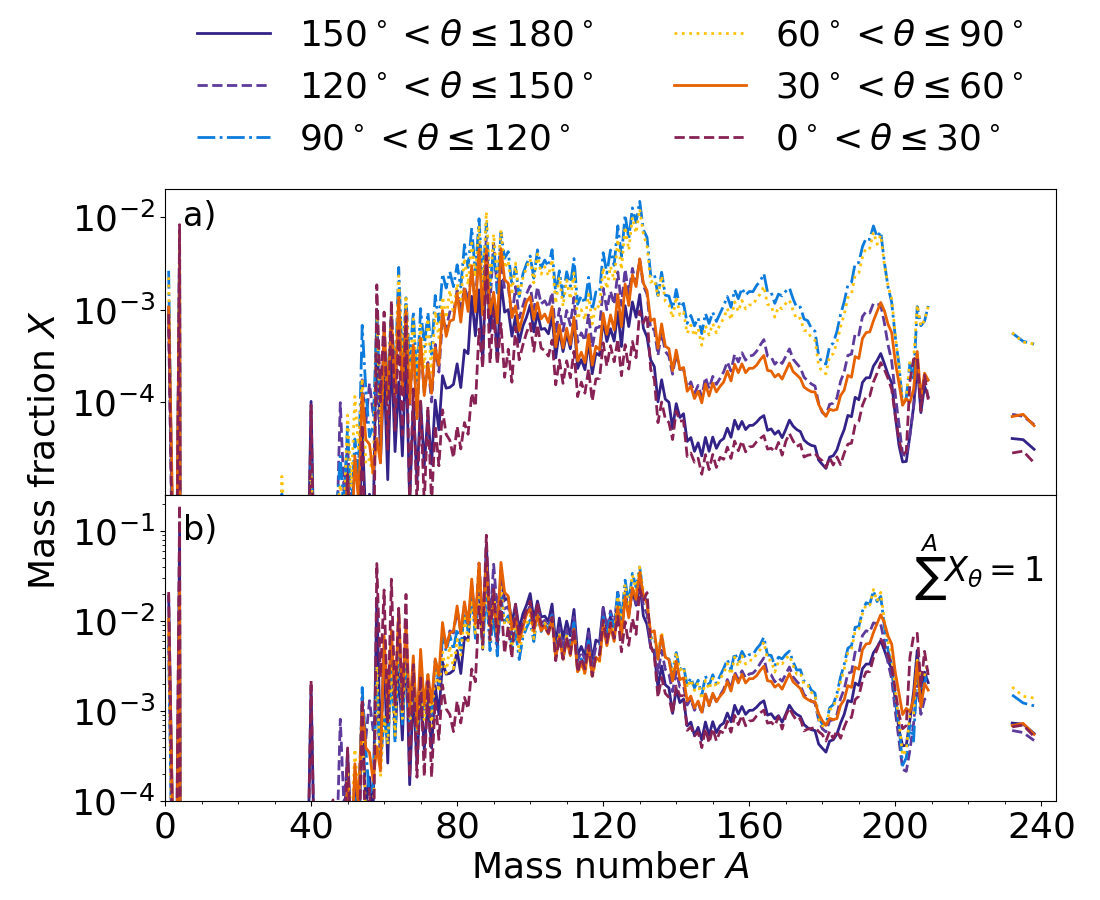}
  \caption{(Color online).  Final mass fractions of the material ejected as a function of the atomic mass $A$
  divided into six angle bins for the DD2-135135 model.  a) Shows the ejected mass fraction $X_\theta$ 
   relative to the total ejected mass, while in b) the mass fractions are renormalised so that the sum over the atomic
     mass numbers are equal to one for each angle bin.}
\label{fig_rpro_aa_angs}
\end{figure}

In Fig.~\ref{fig_rpro_aa_angs} the composition of the ejecta in the polar,  middle and equatorial regions are shown as a function of mass number for the DD2-135135 model.  Fig.~\ref{fig_rpro_aa_angs}a emphasises the differences in ejected yields in the separate angle bins,  with a particular small contribution from the polar regions. Fig.~\ref{fig_rpro_aa_angs}b compares the nucleosynthesis results of the angle regions renormalized such that in that region $\sum_A X_A=1$, showing that the production of $A>140$ nuclei are proportionally smaller in the polar regions compared to the other angle regions.
As can be seen in Fig.~\ref{fig_yedist_ang}, the material in the middle and polar regions has a significantly higher $Y_e$ and is therefore expected to yield a weaker r-process.  However, since the material ejected in both the polar  and middle regions is less massive (see Table~\ref{tab:prop}),  the global effect of the high-$Y_e$ trajectories remains small. In the case of the symmetric SFHo system, a more isotropic distribution of $Y_e$ is found (Fig.~\ref{fig_yedist_ang}), so that a rather similar pattern is observed in all directions. In this case, like in the others, the mass ejected along the poles remains, however, rather low, {\it i.e.} smaller than typically 10\% of the total ejected mass (Table~\ref{tab:prop}).

\begin{figure}
\includegraphics[width=\columnwidth]{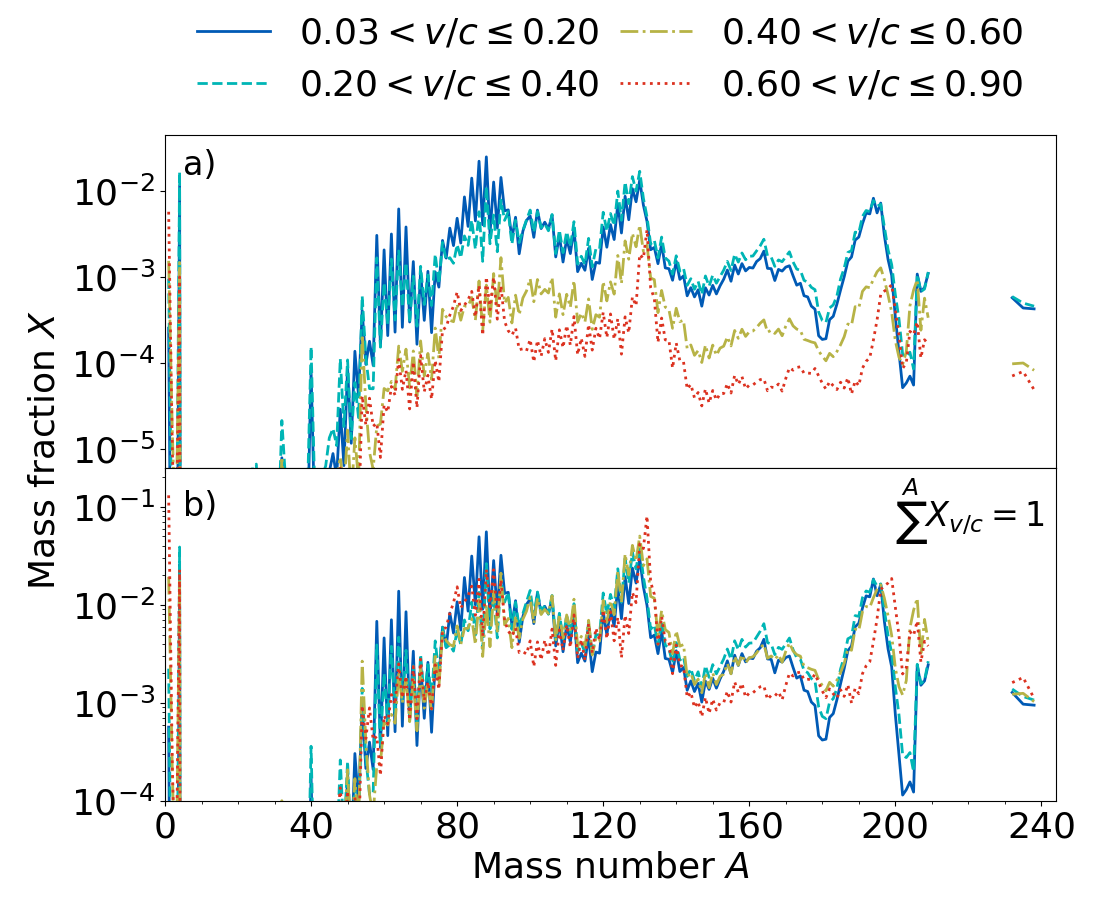}
  \caption{(Color online).  Same as Fig.~\ref{fig_rpro_aa_angs} for four velocity bins. }
\label{fig_rpro_aa_vc}
\end{figure}

The velocity dependence of the isotopic composition  is given in Fig.~\ref{fig_rpro_aa_vc} where panel a) shows the yields relative to the total ejecta mass, and b)  the renormalized mass fractions, so that the sum over the atomic mass for each velocity bin is 1.  The mass elements with the highest velocities ($v\ge 0.6c$) have a small contribution to the overall production of all isotopes.  Comparing only the shape of the abundance distributions in Fig.~\ref{fig_rpro_aa_vc}b, those $v/c>0.6$ trajectories have their second and third r-process peaks shifted to higher mass numbers and a relatively low content in lanthanides with respect to the slow ejecta. This specific distribution is related to the fact that for the fast ejecta not all neutrons are captured and the final composition results from the competition between neutron captures and $\beta$-decays during the fast expansion \citep{goriely2016b}.

\begin{figure}
\includegraphics[width=\columnwidth]{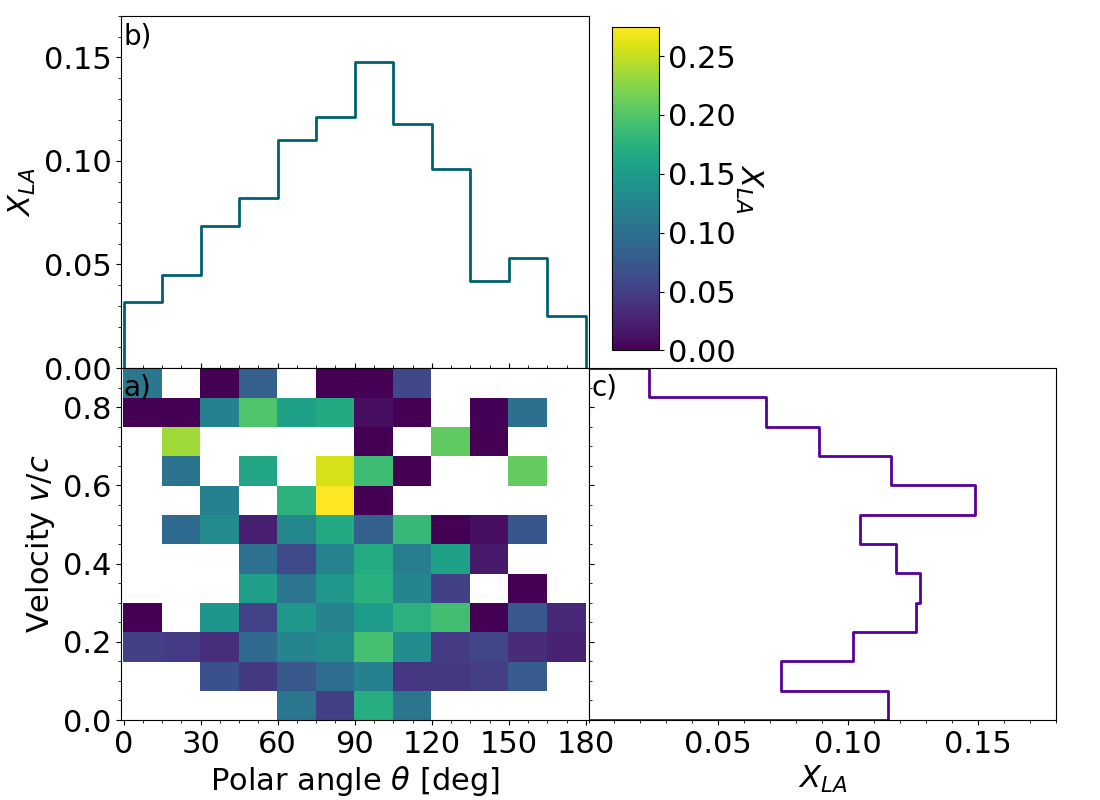}
  \caption{(Color online).  Same as Fig.~\ref{fig_dM2d} for the ejected mass fraction of lanthanides and actinides $X_{LA}$. 
  }
\label{fig_Xla2d}
\end{figure}

The distribution of ejected lanthanides and actinides, $X_{LA}$,  in the $\theta-v/c$ plane is presented in Fig.~\ref{fig_Xla2d}a for the DD2-135135 model, where b) and c) shows the projection of the average $X_{LA}$ fraction on the $\theta$- and $v/c$ axes, respectively.  Both Figs.~\ref{fig_Xla2d}b and \ref{fig_rpro_aa_angs}a show that the production of lanthanides and actinides are the largest in the equatorial plane (see also Table~\ref{tab:prop}). However, as displayed in Fig.~\ref{fig_Xla2d}c, the production of $X_{LA}$ does not seem to have a large dependence on the velocity for $v/c<0.6$ while the production falls rapidly for the fastest escaping mass elements, as mentioned above.

\begin{figure}
\includegraphics[width=\columnwidth]{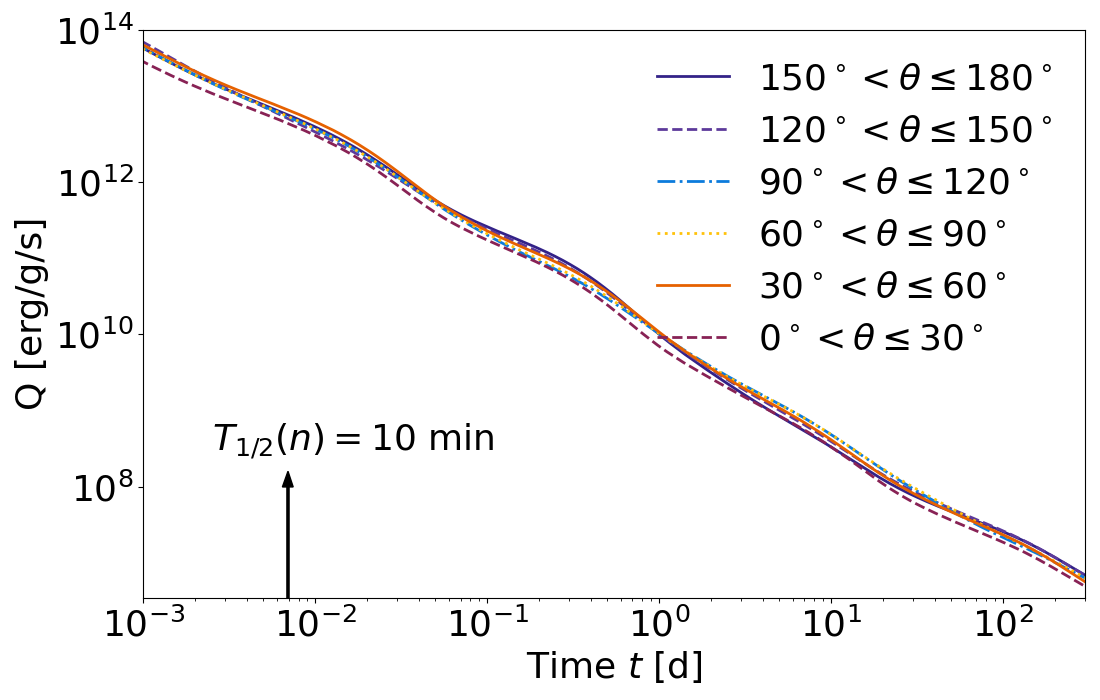}
  \caption{(Color online). Same as Fig.~\ref{fig_Q_t} for the heating rate for the symmetric DD2 model in the six angle bins.   }
\label{fig_Q_t_angles}
\end{figure}

\begin{figure}
\includegraphics[width=\columnwidth]{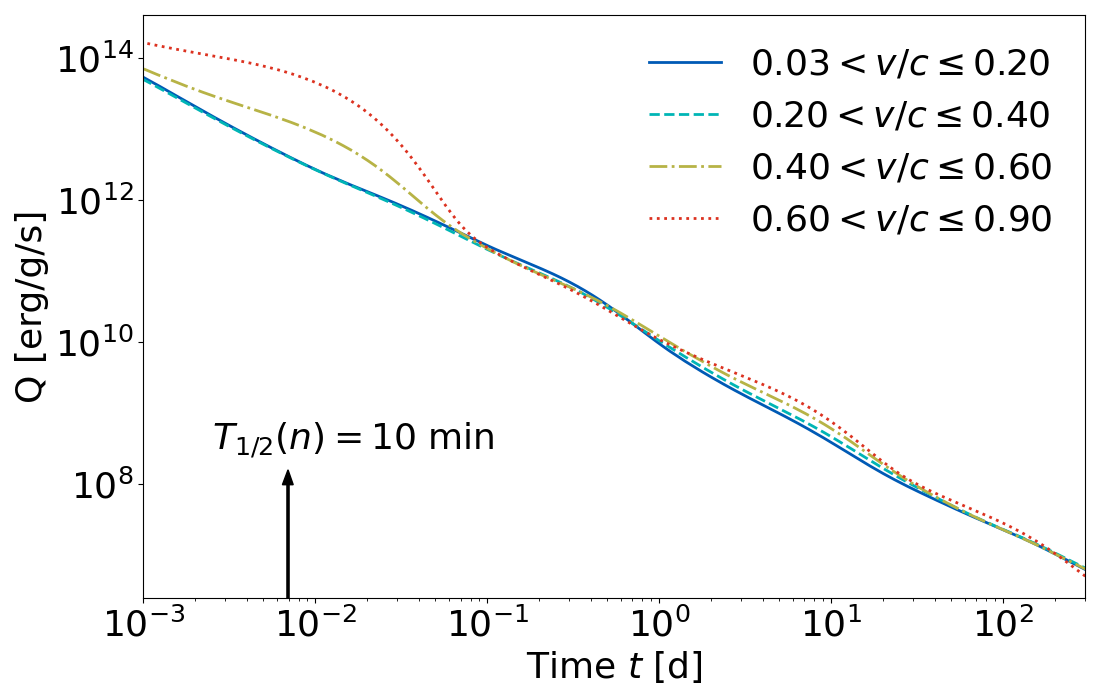}
  \caption{(Color online). Same as Fig.~\ref{fig_Q_t_angles} for the four velocity bins of the DD2-135135 model.  }
\label{fig_Q_t_vc}
\end{figure}

The time evolution of the heating rate in six different angle bins are displayed in Fig.~\ref{fig_Q_t_angles} for the DD2-135135 model. The shape of the curves follows more or less the same $t^{-1/3}$ dependence and are rather similar for the different viewing angles. Some variations are found for the polar directions due to a relatively different composition as shown in Fig.~\ref{fig_rpro_aa_angs}.
On the contrary,  the velocity dependence of the heating rate displayed in Fig.~\ref{fig_Q_t_vc} differs for the highest escape velocities, especially at specific times.  In particular, for the high-velocity ejecta ($v/c>0.6$), the decay of free neutrons at $t\sim 10$~min creates a large bump in the heating rate, and the decay of largely produced $^{132}$Te and $^{132}$I (see Fig.~\ref{fig_rpro_aa_vc}) is responsible for the smaller bump at $t\simeq 10$~d.
In Fig.~\ref{fig_rpro_aa_vc}b, a relative enhancement in the actinide production is also observed for the largest velocity bin ($v/c>0.6$), compared to the low velocities, which leads to an increase in the heating rate for $t \ga 30$~d due to the decay of $^{254}$Cf. 
To a smaller extent, the same features as for the high-velocity ejecta can be seen in the second largest velocity bin ($0.4 < v/c<0.6$), and are explained by the same effects. A rather similar evolution of the heating rate is found for the SFHo-135135 model, and the asymmetric models. We note that the high-velocity tail is discussed as a possible source of the late-time synchrotron emission that was recently observed in GW170817 \citep{hajela2021,nedora2021}.

\section{Comparison to other studies}
\label{sect_comp}

In recent years a variety of publications have focused on the dynamically ejected material from NS-NS mergers \citep[e.g.][]{wanajo2014, palenzuela2015,lehner2016,sekiguchi2015,foucart2016,foucart2020,bovard2017,radice2018a,nedora2021a,martin2018}. However, only a subset of the simulations available in the literature includes neutrino absorption and in addition performs r-process nucleosynthesis network calculations. In the following, we highlight the most relevant differences between our simulations and those of others that also use approximately the same NS mass configurations and EoSs as we do.

The works of \citet{radice2018a} and more recently \citet{nedora2021a} perform fully general relativistic hydrodynamical calculations with neutrino absorption.
Their calculations are performed with the grid-based, adaptive mesh code WhiskyTHC \citep{radice2014} in contrast to our simulations based on the Smoothed Particle Hydrodynamics (SPH) code \citep{oechslin2007,bauswein2013} described in Sect.~\ref{sect_weak}. 
In the optically thick regions they also implement a leakage scheme for the weak interactions, but, in contrast to our ILEAS scheme, without ensuring that the neutrino fluxes obey the diffusion-law and without conserving lepton number in neutrino-trapping regions. In order to describe net neutrino absorption in optically thin regions, they evolve free-streaming neutrinos using a one-moment ("M0") scheme, whereas our ILEAS scheme employs a ray-tracing algorithm. Lacking, to our knowledge, comparisons with reference solutions or with other approximate schemes, we are unable to assess the accuracy of their M0 scheme. We note that while ILEAS has not been benchmarked by Boltzmann solutions in a binary NS merger environment, it showed excellent agreement with accurate neutrino transport solutions for the case of a proto-neutron star formed in a core-collapse supernova \citep{ardevol-pulpillo2019}.

For the r-process nucleosynthesis, \citet{radice2018a} and \citet{nedora2021a} use a mapping technique to calculate the ejecta composition, {\it i.e.} the expansion history of every trajectory is parametrized depending on its density, entropy, electron fraction and velocity at a fixed instant of time (when crossing the sphere of 443~km). On this basis the composition is deduced from the corresponding pre-computed r-process calculation of \citet{lippuner2015}.  In our simulations we perform one r-process calculation per unbound SPH mass element, following its detailed expansion history. 
A subset of models by \citet{radice2018a} as well as \citet{nedora2021a} employs general-relativistic large eddy simulations (GRLES) calibrated to mimic viscosity due to magnetohydrodynamic turbulence. Since the GRLES simulations of \citet{radice2018a} do not include neutrino absorption, we only compare to the calculations without GRLES in the following.  
Note that some complementary information about models discussed in \citet{radice2018a} is provided in  \citet{perego2017}.

Numerical simulations in full general relativity have also been performed by \citet{sekiguchi2015,sekiguchi2016} with a leakage-plus-M1 scheme for neutrino cooling and absorption using finite-volume methods on a nested grid. The corresponding nucleosynthesis calculations are reported in \citet{wanajo2014}. 
When calculating the r-process composition, \citet{wanajo2014} only include ejecta from the orbital plane and choose a representative particle from each $Y_e$-bin from 0.09 to 0.44 giving in total 35 particles.

The recent study by \citet{foucart2020} applied a sophisticated Monte-Carlo scheme for neutrino transport to directly solve the transport equations in low-density regions for a 1.27 -- 1.58~\Msun\ NS-NS merger system. The simulations ran for 5~ms on a Cartesian grid, using high-order finite volume shock-capturing methods with the general relativistic SpEC code, and did not include nucleosynthesis calculations. We note here that the total mass of the binary system in \citet{foucart2020} is 0.15~\Msun\ larger than the total mass of the models discussed above, and in addition a different mass configuration was used. 

\begin{table}
\centering
\caption{Summary of the compared ejecta properties from \citet{radice2018a}, \citet{nedora2021a}, \citet{sekiguchi2015}, \citet{sekiguchi2016}, \citet{foucart2020}, abbreviated as R18, N21, S15, S16, F20, respectively, and this work. All systems have the same total mass $M_1+M_2=2.7$\Msun, except for F20 where it is 0.15~\Msun\ larger. The symmetric mergers have $q=M_1/M_2=1$ for the NSs star masses $M_1$ and $M_2$, while the asymmetric have $q=0.86$, except for F20 where $q=0.80$. }
\begin{tabular}{llcccc}
\hline \hline 
 Ref. & EoS & $q$ & $M_{\rm ej}$         &  $M_{\rm ej}^{v \ge 0.6c}$ & $\langle Y_e\rangle$ \\ 
       &  &         & [$10^{-3}$ \Msun] & [$10^{-5}$ \Msun] &  \\ 
\hline
R18 & DD2 & 1 & 1.4 & 0.2 & 0.23 \\ 
N21 & DD2 & 1 & 1.1 & - & 0.25 \\ 
S15 & DD2 & 1 & 2.1 & - & 0.29 \\ 
This work & DD2 & 1 & 2.0 & 8.7 & 0.27 \\ 
 \hline
 R18 & SFHo & 1 & 4.2 & 3.3 & 0.22 \\ 
 N21 & SFHo & 1 & 2.8 & - & 0.23 \\ 
S15 & SFHo & 1 & 11 & - & 0.31 \\ 
This work & SFHo & 1 & 3.3 & 15.3 & 0.26\\ 
 \hline
F20 & DD2 & 0.80 & 8.3 & - & 0.13 \\
S16 & DD2 & 0.86 & 5 & - & 0.20 \\
This work & DD2 & 0.86 & 3.2 & 17.7 & 0.22\\ 
 \hline
S16 & SFHo & 0.86 & 11 & - & 0.18 \\ 
This work & SFHo & 0.86 & 8.7 & 25.6 & 0.24\\ 
\hline \hline 
\end{tabular} 
\label{tab:comp}
\end{table}

As is common in grid-based codes, the NSs in the simulations of \citet{radice2018a,nedora2021a,sekiguchi2015,foucart2020} are placed in a low-density artificial atmosphere. If not checked carefully,  the atmosphere material may suppress mass ejection or decelerate parts of the ejecta \citep{sekiguchi2016}. This problem is not present in SPH-based simulations where no artificial atmosphere needs to be used. In general, it is assuring that very different numerical schemes, i.e. particle-based and grid-based hydrodynamics methods, yield ejecta properties in the same ballpark noting that, in addition, we employ the conformal flatness approximation, whereas the aforementioned studies consider a fully relativistic evolution. Note that the differences due to the use of the conformal flatness condition are likely to be small, since calculations employing this approximation reproduce post-merger gravitational-wave frequencies of full-general relativity simulations quantitatively very well, implying that our simulations capture well the gravitational field and dynamics of the bulk matter ~\citep[e.g.][]{bauswein2012b}. Also compared to the range of literature data on, for instance, total ejecta masses, our simulations do not seem to feature some systematic offset from fully relativistic simulations.
We emphasize that direct quantitative comparisons with other works remain difficult and should be taken with a grain of salt, as discussed below, in particular when dealing with the total ejected mass. 

We first consider the ejecta masses summarised in Table~\ref{tab:comp}. 
When comparing our results to previously published studies, it should be kept in mind, as already discussed in Sect.~\ref{sect_weak}, that estimating the exact amount of ejected mass remains a difficult task, since it depends on the extraction time as well as on the criterion adopted to count gravitationally unbound material and the radius or volume at which this criterion is evaluated. While we use the $\varepsilon_{\rm stationary}>0$ criterion (see Sect. ~\ref{sect_weak}), \citet{radice2018a} and  \citet{sekiguchi2015} use a more restrictive so-called geodesic criterion which defines a mass element as unbound, if its kinetic energy only is larger than the gravitational one (or, more precisely, if the time component of the fluid four velocity $u_t$ is smaller than $-1$). 
In this case, we expect the ejecta to be less massive than determined by applying our criterion.  
The way how this criterion is applied also differs between published
works. 
While we apply it at a given time $t_{\rm ej}$ in a given volume (with $r> r_{\rm ej}=100$~km), \citet{radice2018a} applies the criterion at a sphere of a given radius (443~km) and time integrates the fluxes at this radius to obtain the ejecta mass. 
\citet{sekiguchi2015,sekiguchi2016} followed our approach, except that
they use $r_{\rm ej}=0$\footnote{One can see in Fig.~\ref{fig_mejec}, however, that our ejecta masses would be very similar if we had used $r_{\rm ej}=0$.} and ran their simulations for a longer time.
\citet{foucart2020} applies a specific version of the Bernoulli criterion to identify unbound ejecta, namely version B of \citet{Foucart2021}, where a variety of ejection criteria are compared and discussed. 
(Note that most merger simulations ---in line with our work--- have not followed the potential ejecta to very large distances, in which case their gravitationally unbound state could not be unambiguously diagnosed.)  For all these reasons, the comparison of the quoted ejected masses should be taken with care. 

This being said, our dynamically ejected mass for the symmetric 1.35--1.35\Msun\ system with the DD2 EoS is found to be 40-80\% larger than the one obtained by \citet{radice2018a} and recently \citet{nedora2021a}, while, with the SFHo EoS, it is  20\% lower than the values found by \citet{radice2018a} and up to 20\% larger than those of \citet{nedora2021a}.  Compared to the results of \citet{sekiguchi2015},  our ejected mass is found to be similar for the DD2 EoS, but 3 times lower for the SFHo EoS. 
When comparing the symmetric and asymmetric merger simulations with the SFHo EoS in \citet{sekiguchi2016}, they yield the same dynamical ejecta mass, while we observe about three times more ejected mass in our asymmetric case.  For the DD2 EoS, their ejected mass is over two times larger for the asymmetric merger,  while our ejected mass only increases by 60\%. For the angular distribution of the ejected mass, \citet{perego2017} (see their Fig.~1) report a $\sin^2\theta$ dependence for the SFHo symmetric model including neutrino heating,  which corresponds well with our SFHo EoS results (see Fig.~\ref{fig_dM2d} for the DD2 EoS).
The seemingly large ejected mass of \citet{foucart2020} for the asymmetric DD2 EoS could also be due to the larger total mass or the smaller mass ratio $q$ of their merger system.

Looking now at the mean electron fractions (cf. Table~\ref{tab:comp} and Fig.~\ref{fig_yedist}), \citet{radice2018a} and \citet{nedora2021a} found with their M0 scheme for the symmetric systems using the DD2 (SFHo) EoSs,  values that are lower compared to our mean $Y_e$ by by 0.04 and 0.02 (0.04 and 0.03), respectively, and \citet{sekiguchi2015} with their leakage-plus-M1 scheme found larger values, namely by 0.02 (0.05). 
However, the $Y_e$ distributions in \citet{radice2018a}, \citet{nedora2021a}, \citet{sekiguchi2015}  and our in Fig.~\ref{fig_yedist} are quite similar since they all show a wide $Y_e$ range spanning from about 0.4 or 0.5 down to at least 0.05. 
In \citet{sekiguchi2016} they also report the $Y_e$ values for asymmetric mergers and find a decrease of the average value relative to symmetric mergers by 0.09 (0.13) compared to 0.05 (0.02) in our case for the DD2 (SFHo) EoS. 
The mean $Y_e$ of 0.13 reported by \citet{foucart2020} for their asymmetric DD2 system is relatively low compared to the other simulations discussed here. While the $Y_e$ distribution of \citet{foucart2020} spans a similar range as in Fig.~\ref{fig_yedist}a,  their distribution peaks at the lowest $Y_e$-value around $\sim 0.05$ (compared to $\sim 0.2$ in our calculations). It remains, however,  difficult to draw conclusions here on the impact of the neutrino treatment since the discrepancy between their $\langle Y_e \rangle$ and ours can also be due  to the different NS masses and, in particular, their lower mass ratio leading to a larger fraction of low-$Y_e$ tidal ejecta. 
As can be observed in Fig.~1 of \citet{perego2017} for the SFHo symmetric merger, the hydrodynamical simulations of \citet{radice2018a} including neutrino interactions produce larger $Y_e$ values in the polar regions compared to the equatorial plane, which can also be seen in our Fig.~\ref{fig_yedist_ang}. However,  as pointed out in Sect.~\ref{sec_ang_vc} (see Fig.~\ref{fig_ye2d}), the mass of the polar ejecta is smaller than the equatorial ejecta, which is also what \citet{perego2017} report. 

When comparing our abundance distribution in Fig.~\ref{fig_rpro_aa} with the simulation of \citet{radice2018a} (see their Fig.~20), some differences can be observed. 
In particular, a different abundance distribution is obtained for $A>90$ nuclei that probably finds its origin in the different nuclear data used as input to the nuclear reaction network (e.g. fission fragment distributions) adopted to estimate the nucleosynthesis yields. However, our isotopic composition is characterised by a systematically smaller production of $A \le 80$ nuclei despite our higher mean electron fraction. The DD2-135135 (SFHo-135135) model is characterised by $\langle Y_e\rangle$=0.27 (0.26) versus 0.23 (0.22) for the equivalent model in \citet{radice2018a}. It remains unclear whether this discrepancy is connected to the nuclear network solver, or the electron fractions, or other trajectory properties that may be differently predicted by \citet{radice2018a}. The mapping technique applied by \citet{radice2018a} and \citet{nedora2021a} to calculate the ejecta nucleosynthesis and taken from the parametrised r-process calculation of \citet{lippuner2015} is also likely to contribute to such differences. The updated calculations of \citet{nedora2021a} show similar nucleosynthesis results as \citet{radice2018a}.

The nucleosynthesis results reported in \citet{wanajo2014} should be compared to our equatorial composition for the SFHo EoS. As can be seen in Fig.~\ref{fig_rpro_aa_angs} and Fig.~\ref{fig_rpro_aa}, the equatorial ejecta dominate the shape of the final abundance distribution for the DD2 EoS, and the same is also true for the SFHo EoS.  Similar to our results, \citet{wanajo2014} report a distribution covering the range of mass numbers from 240 down to $\sim 80$, which is a consequence of the wide $Y_e$ distribution when including neutrino absorption.  \citet{wanajo2014} predict a larger production of $A\sim 80-120$ nuclei compared to our composition, but follow the solar r-process distribution fairly well for $A>120$. Differences in the predicted composition may also originate from the use of different $\beta$-decay rates and fission fragment distributions~\citep{goriely2015,Lemaitre2021}.

As discussed in Sect.~\ref{sec_ang_vc}, the high-velocity ($v/c>0.6$) tail in our simulations contains only a minor fraction of the total ejected mass. Compared to some of our previous simulations \citep{bauswein2013} ignoring neutrino emission and absorption, the inclusion of neutrinos does not give rise to an increase of the total ejecta mass, but it does enhance the mass of ejecta with $v/c>0.6$ by a factor of about two. 
For comparison, \citet{radice2018a} (see their Table 2) find that simulations with the M0 scheme compared to a leakage scheme not accounting for neutrino absorptions tend to increase the total ejected mass as well as the mass of the fast ejecta significantly. Still, in the case of the 1.35-1.35\Msun\ models, they predict masses of fast ejecta with $M_{\rm ej}^{v \ge 0.6c}=1.7\times 10^{-6}$ and $3.3\times 10^{-5}$\Msun\ for the DD2 and SFHo EoSs, respectively, which are about 50 and 5 times lower than those given in Table~\ref{tab:comp}. Similarly, the recent calculations of \citet{nedora2021} (see their Table 1) give masses about 90 and 5 times lower than our results for the fast ejecta of the DD2 and SFHo EoSs, respectively.
These difference are likely to be connected to limitations in the affordable numerical resolution in both our SPH models as well as the grid-based models of \citet{radice2018a}, such that details of the numerical methods have a strong impact. This includes also a possible influence of the aforementioned atmosphere treatment in grid-based codes as well as potentially the way how ejecta properties are extracted, {\it i.e.} on a fixed sphere in \citet{radice2018a} compared to Lagrangian tracers in SPH. Future, better resolved, simulations will have to reduce this uncertainty.

\section{Conclusions}
\label{sect_concl}

In the present work we have studied r-process nucleosynthesis and the subsequent radioactive heating rate produced by material dynamically ejected from four NS-NS merger models including a proper description of neutrino interactions. 
In contrast to more ``conventional'' leakage schemes,  the ILEAS framework coupled to the hydrodynamical calculations follows the detailed evolution of the electron fraction and its changes due to weak interactions through a neutrino-equilibration treatment in the high-density regions and an absorption module in semitransparent conditions \citep{ardevol-pulpillo2019}.  This allows us to study the impact of the neutrino interactions with nucleons on the composition of the dynamical ejecta of NS mergers.
Our four merger models include two EoSs, namely DD2 and SFHo, and for each EoS we have included symmetric and asymmetric merger cases, {\it i.e.}  systems of 1.35-1.35\Msun\ and 1.25-1.45\Msun\ NSs, which all lead to the formation of a longer-lived neutron star surrounded by an accretion torus as remnant.
The second part of this work \citep{just2021} presents kilonova light curve calculations that follow from the nucleosynthesis results presented here.
%
%
%
%
%
%
%
%

The main conclusions from our study can be summarised as follows:
\begin{itemize}
\item A proper inclusion of neutrino interactions yields dynamical ejecta with a solar-like composition down to $A \simeq 90$ compared to 140 without neutrino interactions. This includes in particular a significant enrichment in Sr. This conclusion is rather similar to the one drawn initially by \citet{wanajo2014}, though the final abundance distribution obtained in our case with different nuclear physics inputs is found to be in much closer agreement with the solar system distribution. 

\item The composition of the ejected matter as well as the corresponding heating rate are found to be rather independent of the system mass asymmetry and the EoS. In the four hydrodynamical models, despite differences in the initial electron fraction distributions, rather similar nucleosynthesis results are found.   This approximate degeneracy in abundance pattern and heating rates can be favourable for extracting the ejecta properties (such as masses, opacities, expansion velocities) from kilonova observations.  
However, our study has two clear limitations. First, only the dynamical ejecta are taken into account. The hypermassive NS or BH-torus wind ejecta may not follow the same trend with EoS and mass ratio as the dynamical ejecta. Second, here we only consider delayed-collapse cases in which a NS remnant survives for a longer period of time after the merger. Cases of prompt BH formation may exhibit different nucleosynthesis signatures, albeit the ejecta in these cases are typically much less massive \citep[e.g.][]{bauswein2013}. 

\item Uncertainties in the initial electron fraction impact not only the composition and in particular the amount of lanthanides and actinides ejected, but also the heating rate, mainly  10~min, 7~h or 30~d after merger through the $\beta$-decay of residual nuclei. Spontaneous fission of $^{254}$Cf or $A=222-225$ $\alpha$-decay chains contribute significantly less to the heating rate when weak nucleonic processes are taken into account.

\item  The nucleonic weak processes have the largest impact on the composition of the material ejected in the polar regions but also affect the r-process efficiency in the equatorial plane.

\item Even with nucleonic weak processes, the fast expanding layers are found to include a non-negligible amount of free neutrons that should leave an observable signature in the light curve and may contribute to late time X-ray emission \citep{hajela2021}.




\end{itemize}

In addition to the important role of  nucleonic $\beta$-processes on the final composition of the ejecta, the final abundance distribution as well as decay heat are known to be affected by nuclear uncertainties. Despite numerous works dedicated to the impact of nuclear physics uncertainties in this context, the proper inclusion of  neutrino interactions may change the conclusion of such sensitivity studies. In particular, as already shown by \citet{Lemaitre2021}, the role of fission may be significantly reduced if neutrino interactions are included and found to affect the initial neutron richness. 
Finally, before drawing conclusions on the contribution of NS mergers to the Galactic enrichment, the outflow generated during the evolution of the merger remnant must be consistently included in the simulations and the discussion \cite[see][]{just2015}, which will be the objective of a forthcoming paper.  Much work remains to be performed in this respect.

\section*{Acknowledgments}
SG acknowledges financial support from F.R.S.-FNRS (Belgium). This work has been supported by the Fonds de la Recherche Scientifique (FNRS, Belgium) and the Research Foundation Flanders (FWO, Belgium) under the EOS Project nr O022818F. 
 The present research benefited from computational resources made available on the Tier-1 supercomputer of the Fédération Wallonie-Bruxelles, infrastructure funded by the Walloon Region under the grant agreement n$^\circ$1117545 and the Consortium des Équipements de Calcul Intensif (CÉCI), funded by the Fonds de la Recherche Scientifique de Belgique (F.R.S.-FNRS) under Grant No. 2.5020.11 and by the Walloon Region. 
OJ and  AB acknowledge support by the European Research Council (ERC) under the European Union's Horizon 2020 research and innovation programme under grant agreement No. 759253. AB was supported by Deutsche Forschungsgemeinschaft (DFG, German Research Foundation) - Project-ID 279384907 - SFB 1245 and - Project-ID 138713538 - SFB 881 (``The Milky Way System'', subproject A10) and acknowledges the support by the State of Hesse within the Cluster Project ELEMENTS. OJ acknowledges computational support by the HOKUSAI supercomputer at RIKEN.
At Garching, funding by the European Research Council through Grant ERC-AdG No.~341157-COCO2CASA and by the Deutsche Forschungsgemeinschaft (DFG, German Research Foundation)
through Sonderforschungsbereich (Collaborative Research Centre) SFB-1258 ``Neutrinos and Dark Matter in Astro- and Particle Physics (NDM)'' and under Germany's Excellence Strategy through
Cluster of Excellence ORIGINS (EXC-2094)---390783311 is acknowledged.

\section*{Data availability}
The data underlying this article will be shared on reasonable request to the corresponding author.

\bibliographystyle{mnras}
\bibliography{MyZoteroLibrary.bib}

\end{document}